\documentclass[a4paper,onecolumn,11pt,accepted=2023-07-07]{quantumarticle}
\pdfoutput=1
\usepackage[english]{babel}
\usepackage[utf8]{inputenc}
\usepackage[numbers,sort&compress]{natbib}
\usepackage{amsthm}
\usepackage{amsmath}

\DeclareMathOperator*{\argmin}{arg\,min}
\usepackage{mathtools}
\usepackage{physics}
\usepackage{xcolor}
\usepackage{graphicx}
\usepackage[left=23mm,right=13mm,top=35mm,columnsep=15pt]{geometry} 

\usepackage{adjustbox}
\usepackage{placeins}
\usepackage[T1]{fontenc}
\usepackage{lipsum}
\usepackage{csquotes}
\usepackage[pdftex, pdftitle={Article}, pdfauthor={Author}]{hyperref}
\usepackage{xcolor}
\hypersetup{
    colorlinks,
    linkcolor={red!50!black},
    citecolor={blue!50!black},
    urlcolor={blue!80!black}
}
\usepackage{booktabs}
\usepackage{multirow}
\usepackage{listings}
\usepackage{xcolor}

\definecolor{codegreen}{rgb}{0,0.6,0}
\definecolor{codegray}{rgb}{0.5,0.5,0.5}
\definecolor{codepurple}{rgb}{0.58,0,0.82}
\definecolor{tqblue}{HTML}{08293d}
\definecolor{backcolour}{HTML}{fefdf5}

\lstdefinestyle{mystyle}{
    backgroundcolor=\color{backcolour},   
    commentstyle=\color{codegreen},
    keywordstyle=\color{magenta},
    numberstyle=\tiny\color{codegray},
    stringstyle=\color{codepurple},
    basicstyle=\ttfamily\footnotesize\color{tqblue},
    breakatwhitespace=false,         
    breaklines=true,
    postbreak=\mbox{\textcolor{magenta}{$\hookrightarrow$}\space},                 
    captionpos=b,                    
    keepspaces=true,                 
    numbers=left,                    
    numbersep=5pt,                  
    showspaces=false,                
    showstringspaces=false,
    showtabs=false,                  
    tabsize=2
}

\newcommand{\lr}[1]{\ensuremath{\left( #1 \right)}}

\usepackage{algorithm}
\usepackage{algpseudocode}

\makeatletter
\newenvironment{heuristic}[1][htb]{%
    \renewcommand{\ALG@name}{Heuristic}
   \begin{algorithm}[#1]%
  }{\end{algorithm}}
\makeatother

\begin{document}
\title{Molecular Quantum Circuit Design: A Graph-Based Approach}
\author{Jakob S. Kottmann}
\affiliation{{Institute of Computer Science, University of Augsburg, Germany }}
\orcid{0000-0002-2445-2701}
\email{jakob.kottmann@uni-a.de}
\maketitle

\begin{abstract}
Science is rich in abstract concepts that capture complex processes in astonishingly simple ways. A prominent example is the reduction of molecules to simple graphs.
This work introduces a design principle for parametrized quantum circuits based on chemical graphs, providing a way forward in three major obstacles in quantum circuit design for molecular systems: Operator ordering, parameter initialization and initial state preparation. It allows physical interpretation of each individual component and provides an heuristic to qualitatively estimate the difficulty of preparing ground states for individual instances of molecules.
\end{abstract}

\maketitle

\begin{figure}
    \centering
    \includegraphics[width=0.7\textwidth]{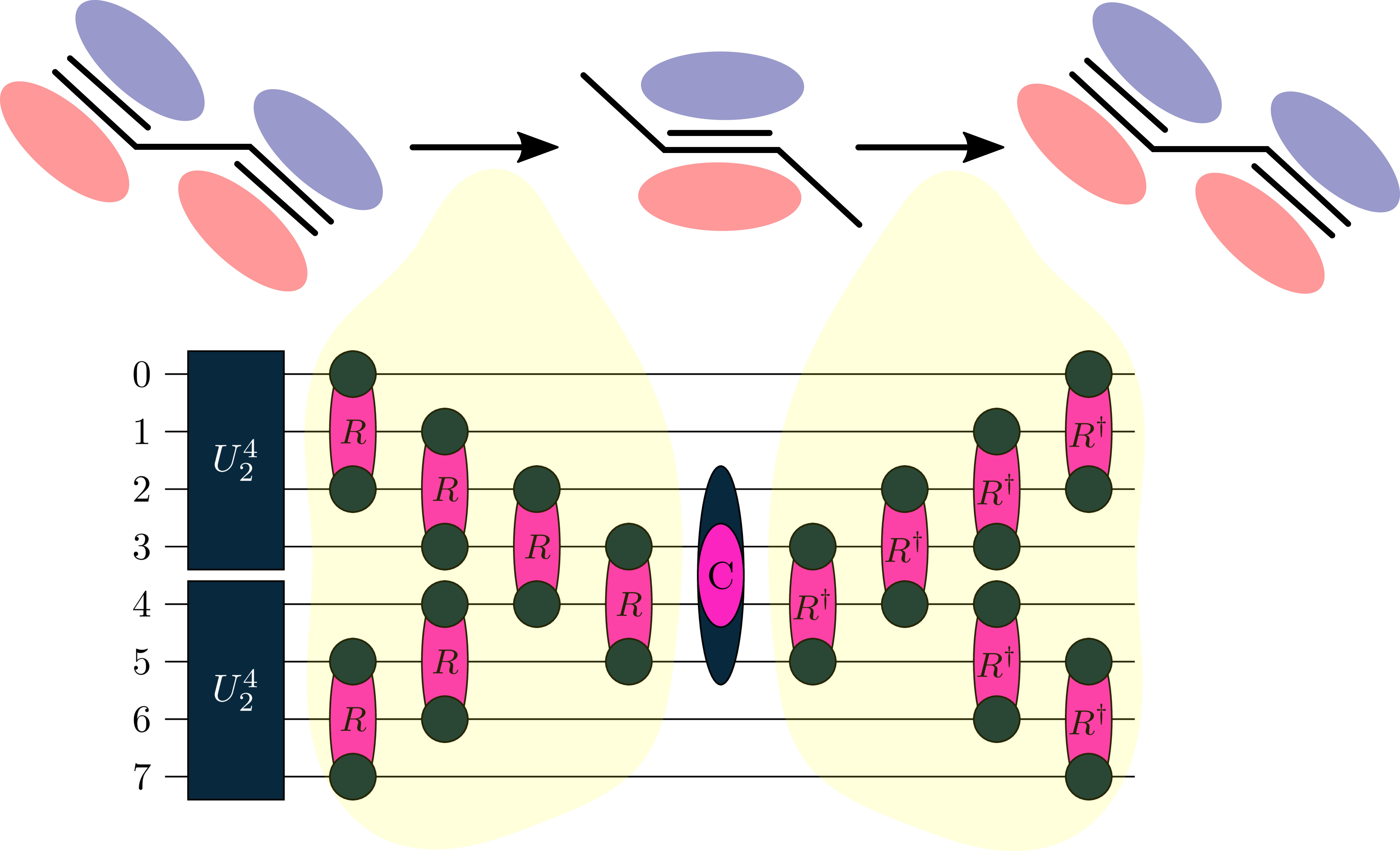} 
    \caption{Illustration of the basic concepts developed in this article connecting chemical graphs with orbitals and quantum circuits. The circuit can be employed to the $\pi$-system of C$_4$H$_4$ (sketched in the Figure) as well as the linear H$_4$ and BeH$_2$ (see the explicit examples in the main text).}
    \label{fig:overview_cartoon}
\end{figure}
\section{Introduction}
An ongoing question in variational quantum approaches~\cite{bharti2022noisy, cerezo2020variational, mcclean2016theory} is how to efficiently construct parametrized quantum circuits. Desired criteria are shallowness, locality in qubit connectivity, expressibility~\cite{sim2019expressibility} with respect to the problem class at hand, as well as well behaved convergence in parameter optimization.\\
Current day quantum computers come with significant device noise and restricted qubit connectivity, making shallowness and locality of the respective circuits a necessity.
This restriction motivated the so called \textit{hardware-efficient} approaches~\cite{kandala2017hardware} were quantum circuits are designed to be most favorable with respect to the available quantum hardware. If used in a problem agnostic fashion, hardware-efficient circuit design usually requires high parameter counts and comes with poor convergence properties. Without a reliable procedure for parameter initialization the optimization becomes furthermore tedious.~\cite{mcclean2018barren}
From this, it can be assumed, that good initialization heuristics, including machine learning approaches~\cite{cervera2021meta, zhang2020collective, sauvage2021flip, ceroni2023generating}, are more likely found for structured quantum circuits. Problem-aware circuit designs, as for example unitary coupled-cluster~\cite{anand2022quantum}, naturally have more structure than purely hardware-efficient designs and typically come with low parametrization and improved convergence at the expense of deep non-local circuits. Due to the increased structure convergence properties are significantly improved but still not reliable across different model systems.\\

Apart from device noise, the construction of efficient circuits is important as, at the end of the day, a shallower circuit with less parameters and improved convergence properties will result in reduced runtime on quantum devices or might even lead to classically tractable simulation methods as in Ref.~\cite{kottmann2022optimized} where a problem-aware but hardware-efficient circuit design was developed.
Pioneering physical motivated circuit designs in electronic structure~\cite{mcclean2016theory, lee2018generalized, barkoutsos2018quantum, wecker2015progress} relied on trotterized time evolutions leading to comparable deep circuits and the need for heuristics on the optimal ordering~\cite{grimsley2019trotterized,  evangelista2019exact, Izmaylov2020order} of the primitive operations.
Adaptive approaches, most prominent in the form of qubit coupled-cluster~\cite{ryabinkin2018qubit, ryabinkin2020iterative}, adapt-vqe~\cite{grimsley2019adaptive, tang2021qubit, grimsley2023adapt} and related selective approaches~\cite{Fedorov2022unitaryselective} tackle the ordering problem through a greedy approach. Together with~\cite{kottmann2022optimized, anselmetti2021local, gard2020efficient}, these approaches often provide a middle-ground between hardware-efficient and physically-motivated design principles and are promising candidates for a path towards fully automatized and reliable circuit construction. 
A rule of thumb is, that the success  of the (adaptive) process is more likely when it starts from a good initial state.~\cite{kottmann2022optimized, rubin2022compressing, kottmann2021feasible, khamoshi2020correlating, khamoshi2022agpbased} 
In a similar fashion, algorithms like the quantum phase estimation~\cite{aspuru2005simulated} or quantum imaginary time evolution~\cite{motta2020determining, sun2021quantum} will profit from improved initial states through increased success probabilities.\\
In this work we will focus on the construction of shallow quantum circuits with reasonably local qubit connectivity using the concept of chemical graphs.  
The most important graph is directly represented by a classically tractable circuit, and for each additional chemical graph the quantum circuit is extended with an $U_\text{R}U_\text{C}U_\text{R}^\dagger$ motif illustrated in Fig.~\ref{fig:overview_cartoon}. Here $U_\text{R}$ rotates the molecular orbitals such that they represent the given chemical graph and $U_\text{C}$ is built from (fermionic) correlators.
The developed design principles provide a way forward in all three major obstacles in quantum circuit design for molecular systems:
\begin{itemize}
    \item Operator ordering: A given chemical graph G gives rise to a specific ordering of fermionic unitary single and double excitations grouped in $U_\text{G}=U_\text{R}U_\text{C}U_\text{R}^\dagger$ motifs. The ordering of the individual structural motifs is given by the relative importance of the graphs which can be estimated through physically-motivated rules.
    \item Parameter Initialization: Angles of double excitations are initialized to zero in order to guarantee an overall energy improvement. Single excitations are initialized based on suitable initial guesses for orbitals representing the chemical graph. This ensures improved convergence in the optimization of the angles.
    \item Initial state: The correspondence between a single chemical graph and the separable pair approximation~\cite{kottmann2022optimized} is exploited to allow for a classically simulable initial state.
\end{itemize}

\section{Molecular Quantum Circuit Design}
In the following we will describe how molecular quantum circuits can be constructed through chemical model systems that we will organize in three layers of abstraction: the chemical graph that represents the molecule of interest through a set of vertices (nuclei) and edges (bonds), the orbitals (one-body wavefunctions in spatial space that are mapped to qubits)
which can be rotated to resemble such a graph, and the quantum circuits build from those rotations combined with two-electron correlators.\\

We will start with two electron systems and demonstrate that an orbital-optimized form of the classically tractable separable pair approximation of Ref.~\cite{kottmann2022optimized} solves those systems exactly.  The corresponding circuits can be transferred to \textit{effective} two-electron systems like the LiH molecule -- also a prominent proof-of-principle example in the literature. The ground states of all those systems can be approximated almost exactly (with energetic errors below the millihartree threshold) with the constructed circuits. We will use this furthermore to introduce and illustrate important concepts like orbital rotations, separable pair approximations and the frozen core approximation in an illustrative way.
In the sections thereafter we will then discuss how these concepts can be extended beyond (effective) two-electron systems.
\begin{figure*}
    \centering
    \includegraphics[width=0.32\textwidth]{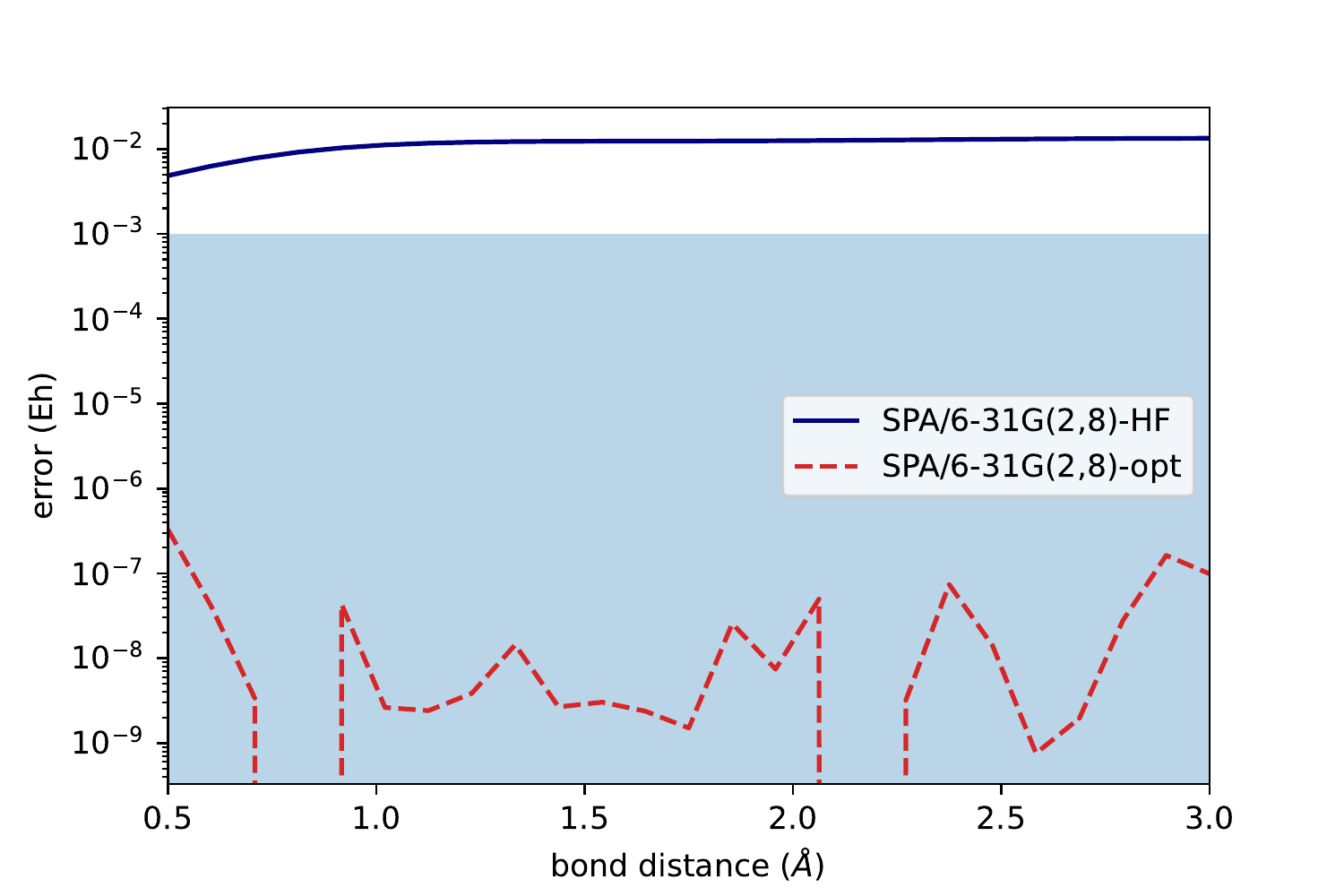}
    \includegraphics[width=0.32\textwidth]{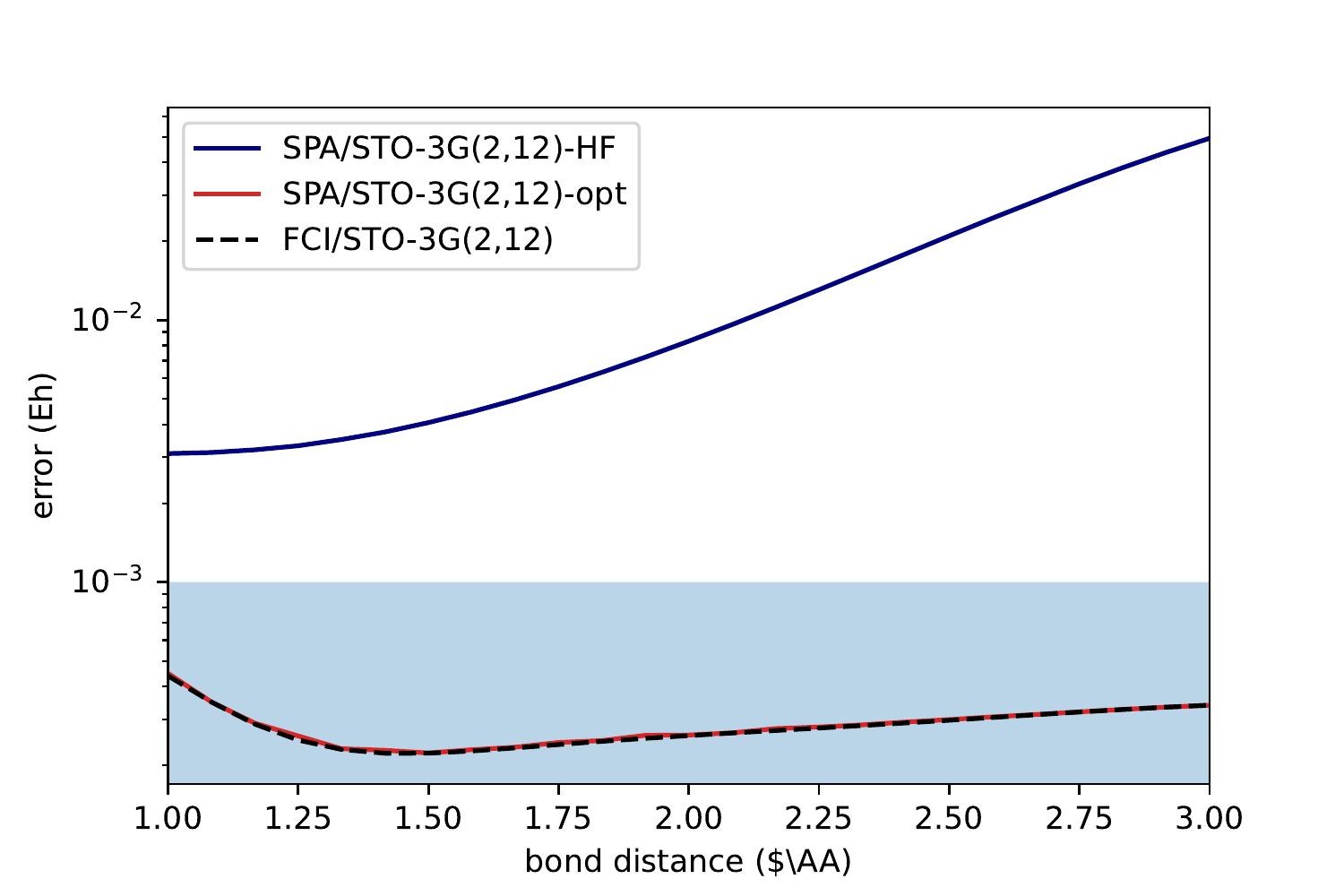}
    \includegraphics[width=0.32\textwidth]{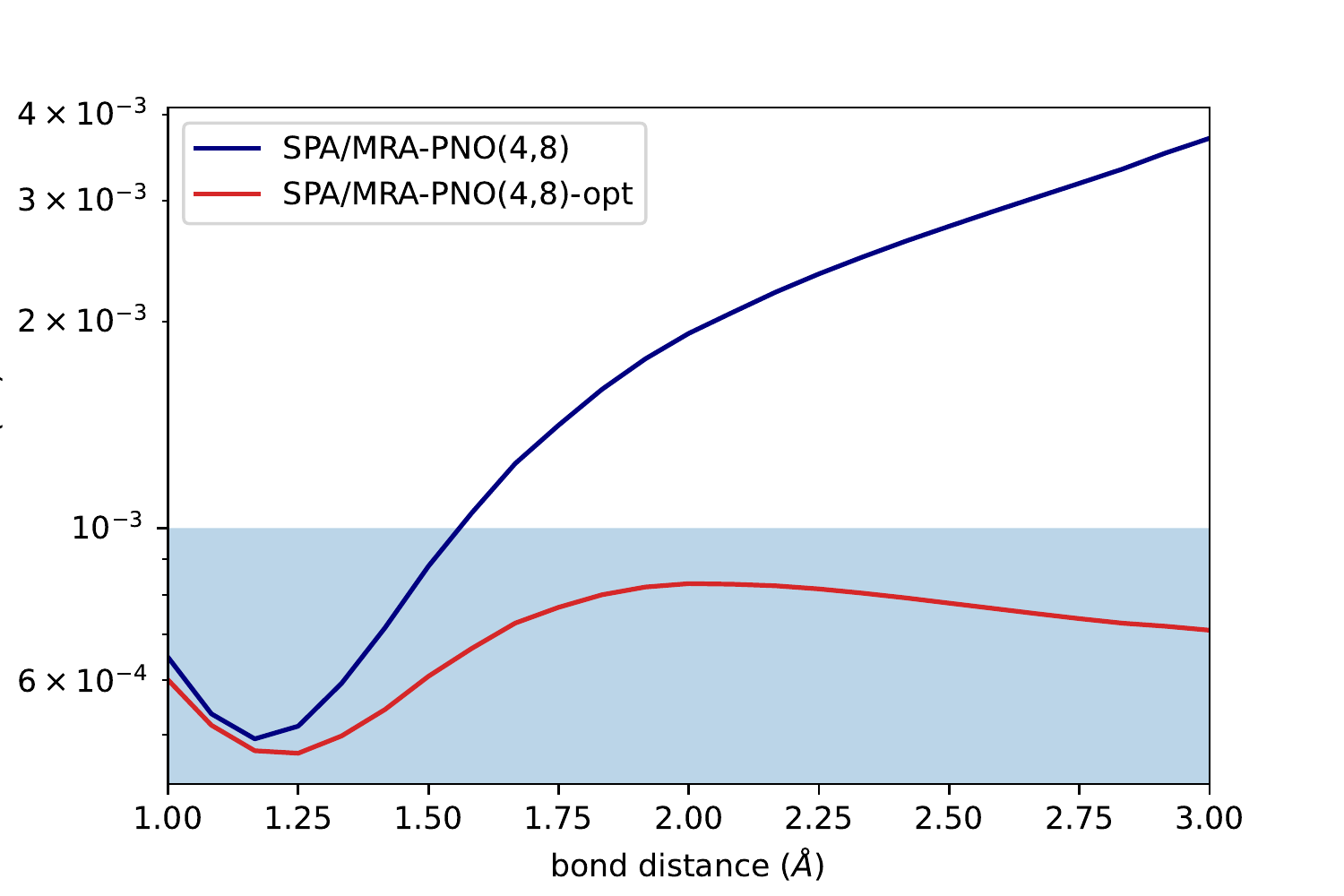}
    \caption{Near-Exactness of separable pair approximations (single edge molecular graphs) for H$_2$ and LiH in different bases. Left: Performance of the $U_2^8$ circuit~\eqref{eq:h2_spa_circuit} for the H$_2$/6-31G(2,8) system using standard Hartree-Fock (HF) or optimized orbitals (opt) with FCI/6-31G(2,8) as reference. Orbital optimization is necessary to reach the exact ground state energy. Center: Performance of the related $U_2^{12}$ circuit for the LiH/STO-3G(2,12) with frozen core orbital and FCI/STO-3G(4,14) as reference, showing that orbital optimization is necessary and the frozen-core error is negligible.  Right: $U_\text{SPA}^{(4,8)}$ circuit (the same as in Eq.~\eqref{eq:h4_spa}) for LiH/MRA-PNO(4,8) with an adaptive basis that ensures significant core correlation and reference energy FCI/MRA-PNO(4,8) showing that the 4 electrons of LiH can be can be separated in a core and valence pair. More details on the LiH results are provided in the appendix. The millihartree accuracy threshold is marked with blue.}
    \label{fig:h2_lih_results}
\end{figure*}
\subsection{Exact and Near-Exact Circuits: (Effective) Two Electron Systems}
The hydrogen molecule was used prominently in the pioneering work on variational quantum algorithms~\cite{peruzzo2014variational} and has since then served as a toy model for a myriad of approaches in variational quantum computing. Here we will use it to introduce important concepts for the following parts of this article and to clarify why it is not an ideal benchmark system for novel algorithmic approaches to quantum chemistry as its eigenstates can be prepared efficiently by a low-depth and local circuit with low variable count that belongs to a class of classically tractable quantum circuits.\\
We will begin with the smallest representation of H$_2$ that still admits correlation beyond a mean-field treatment. Here the electrons are allowed to move within two spatial orbitals, usually represented by linear combinations of atomic orbitals placed on the center of the two hydrogen atoms (see~\ref{sec:molecular_hamiltonian} and~\ref{sec:orbital_rotation_circuits} for more details) . We will represent this graphically as
\begin{align}
    \raisebox{0.09cm}{$\psi_\text{R} \equiv\;$} \includegraphics[width=0.05\textwidth, angle=0]{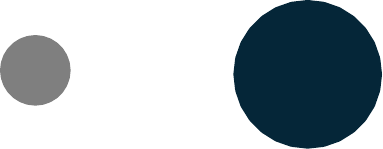},\quad    \raisebox{0.09cm}{$\psi_{\text{L}}\equiv\;$}\includegraphics[width=0.05\textwidth, angle=0]{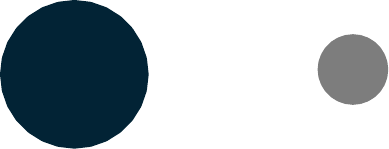}\label{eq:h2_atomics_cartoon}
\end{align}
where gray dots represent the atomic centers and the dark spheres the spherical symmetric atomic orbitals.
A prominent \textit{basis set} that provides those atomic orbitals is the STO-3G set (\textit{S}later \textit{T}ype \textit{O}rbitals made from \textit{3} contracted \textit{G}aussian functions). For more clarity we will denote the number of electrons and spin-orbitals (qubits) in parenthesis, so in this case we have H$_2$/STO-3G(2,4).\\
The standard approach to represent molecules graphically is given by molecular graphs (or Lewis formulas) with atomic nuclei as vertices and \textit{chemical bonds}~\footnote{Note, that this is not an exact definition of a chemical bond, but rather an illustrative graphical simplification predominantly used when discussing chemical properties.} as edges. For H$_2$ this looks like
\begin{align}
    \includegraphics[width=0.1\textwidth]{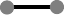}\label{eq:h2_graph}
\end{align}
where the edges are interpreted as electron pairs that connect two atoms. We can prepare a {4-qubit} wavefunction
\begin{align}
    U_2^4\ket{0000} = \cos\lr{\frac{\theta}{2}}\ket{1100} + \sin\lr{\frac{\theta}{2}}\ket{0011},\nonumber
\end{align}
that represents this single bond situation with the circuit {given as}
\begin{align}
    \raisebox{1cm}{$U_2^4 =$}\includegraphics[width=0.2\textwidth]{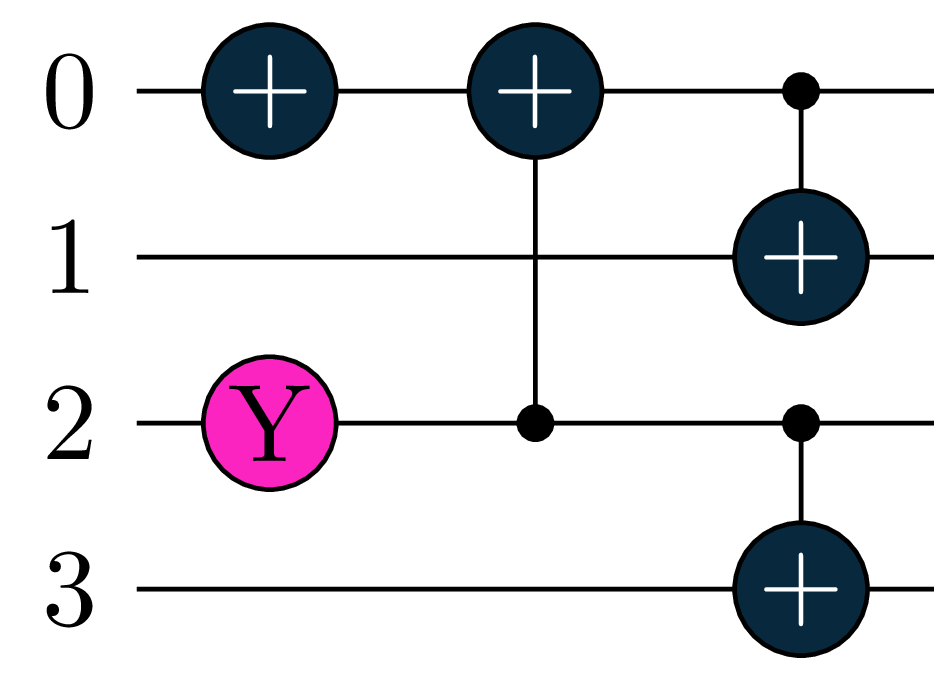}\label{eq:h2_spa_circuit_4}.
\end{align}
The free parameter $\theta$ enters via the parametrized $e^{-i\frac{\theta}{2}\sigma_y}$ single qubit rotation gate in~\eqref{eq:h2_spa_circuit_4}, with the four qubits representing four spin orbitals $\phi_{i_\uparrow}, \phi_{i_\downarrow}$ that are formed by linear combinations of the atomic spin orbitals
\begin{align}
    \phi_{i_\uparrow} = \sum_k c_{ik} \psi_{k_\uparrow}\label{eq:orbital_coefficients}
\end{align}
with $\psi_{k\uparrow}  \in \left\{ \psi_{\text{R}\uparrow}, \psi_{\text{L}\uparrow}\right\}$ (analog for the spin-down orbitals).
{In most applications, including this work, the spin-up and spin-down orbitals are assembled through the same coefficients $c_{i_{\uparrow}k_{\uparrow}}=c_{i_{\downarrow}k_{\downarrow}}=c_{ik}$. For our 4-qubit (2 spatial orbitals) H$_2$ system, we can write Eq.~\eqref{eq:orbital_coefficients} explicitly as matrix equation (omitting the spin label)}
\begin{align}
    \begin{pmatrix}
        \phi_0 \\ \phi_1 
    \end{pmatrix}
    =
    \begin{pmatrix}
        c_{00} & c_{01} \\
        c_{10} & c_{11}
    \end{pmatrix}
    \begin{pmatrix}
        \psi_\text{R} \\ \psi_\text{L} 
    \end{pmatrix}.
\end{align}
{In order to describe the resulting orbitals $\phi_k$, we will resort to the graphical depiction of Eq.~\eqref{eq:h2_atomics_cartoon} and represent the coefficients $c_{ik}$ as small circles placed on the corresponding atomic centers. The size of the circles will correspond to the relative size between the coefficients and the color will indicate sign changes, with red being used for negative coefficients. Lets consider molecular orbitals for H$_2$ where all coefficients are equal in size forming a \textit{bonding} and \textit{anti-bonding} molecular orbital from the atomic basis functions}
\begin{align}
    \frac{1}{\sqrt{2}}\begin{pmatrix}
        1 & 1 \\
        1 & -1
    \end{pmatrix}\; \equiv \;\raisebox{-0.45cm}{\includegraphics[width=0.075\textwidth]{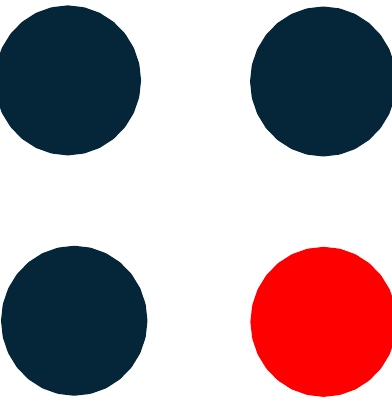}}\;\;\raisebox{-0.35cm}{.}\label{eq:h2_canonical}
\end{align}

Combined with the right choice of orbitals -- {in this case, the canonical orbitals from Eq.~\eqref{eq:h2_canonical}} --  the wavefunction prepared by circuit~~\eqref{eq:h2_spa_circuit_4} describes the 2 orbital (4 qubit) H$_2$ system exactly for all possible bond distances.\\ 

{The orbital rotations in Eq.~\eqref{eq:orbital_coefficients} can be represented as a unitary operation in the second quantized formulation}
\begin{align}
U_{\text{R}_{0}^{1}} = e^{-i\frac{\varphi}{2}G}\; \text{with } G=a^\dagger_{0\uparrow} a_{1\uparrow} + a_{0\downarrow}^\dagger a_{1\downarrow} + h.c.\nonumber
\end{align}
a so-called fermionic basis rotation (see appendix~\ref{sec:orbital_rotation_circuits} for more details).   
Translated to a quantum circuit this orbital rotating unitary looks like
\begin{align}
    \raisebox{1.2cm}{$U_{\text{R}_{0}^{1}} =$}\;
    \includegraphics[width=0.175\textwidth]{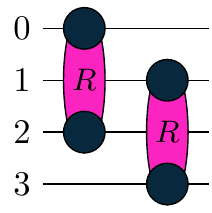}\label{eq:h2_R}
\end{align}
where the two gates represent single-electronic excitations within $R$ and $L$ orbitals (qubits 0 and 2 for spin-up and qubits 1 and 3 for spin-down electrons respectively), each representable with two three-qubit Pauli rotations ($XZY$ and $YZX$ -- see appendix~\ref{sec:orbital_rotation_circuits} for an explicit decomposition).\\
Alternatively the orbitals can be optimized in a more direct way by using a first order expansion of the transformed Hamiltonian $H'=U_\text{R}^\dagger H U_\text{R}$ and optimizing the angles $\varphi$ in the resulting functional. After the optimized angles are found, a new molecular Hamiltonian can be constructed from the optimized orbitals.
In the case of H$_2$/STO-3G(2,4), the optimized orbitals are the symmetric and anti-symmetric combination of the atomic orbitals from Eq.~\eqref{eq:h2_canonical}
\begin{align}
    \includegraphics[width=0.1\textwidth, angle=90]{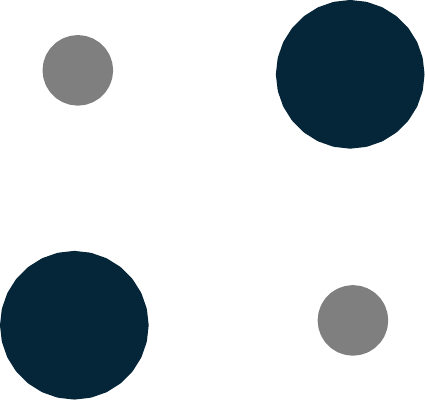}\;\;\raisebox{0.6cm}{$\xrightarrow{U_{\text{R}}}$}\;\;\;
   \includegraphics[width=0.1\textwidth]{h2_hf_orbitals.pdf}\;\;\label{eq:h2_optimized_orbitals_cartoon},
\end{align}
{meaning, adding the circuit~\eqref{eq:h2_R} to circuit~\eqref{eq:h2_spa_circuit_4} and optimizing the angles has the same effect as using only circuit~\eqref{eq:h2_spa_circuit_4} with optimized orbitals. In the following part of this work, we will frequently resort to this graphical representation of the action of from fermionic rotations $U_\text{R}$.}
In this case, the optimal orbitals with respect to the wavefunction prepared by $U_2^4$ are identical to the canonical Hartree-Fock orbitals and their optimal linear combination is the same for all bond distances. This is however not always the case and is due to the high degree of symmetry in the H$_2$/STO-3G(2,4) system.\\

On first glance, it might seem counter-intuitive that the orbitals in~\eqref{eq:h2_optimized_orbitals_cartoon} are optimal even for large bond distances, where the molecule is essentially dissociated. 
In the dissociated limit, the optimal angle in circuit~\eqref{eq:h2_spa_circuit_4} is $\pi/2$ making the generated qubit wavefunction
\begin{align}
    \ket{\Psi}_\text{qubit} \xrightarrow[distance \rightarrow \infty]{bond} \frac{1}{\sqrt{2}}\lr{\ket{1100} - \ket{0011}}.
\end{align}
with the corresponding fermionic expression
\begin{align}
    \ket{\Psi}_\text{f} &= \frac{1}{\sqrt{2}}\left(a_{0\uparrow}^\dagger a_{0\downarrow}^\dagger - a_{1\uparrow}^\dagger a_{1\downarrow}^\dagger\right) \ket{\text{vac}}\\
    & = \frac{1}{\sqrt{2}}\lr{a_{\text{L}\uparrow}a_{\text{R}\downarrow} + a_{\text{R}\uparrow}a_{\text{L}\downarrow}}\ket{\text{vac}}
\end{align}
where the first line denotes the wavefunction with the optimal molecular orbitals~\eqref{eq:h2_optimized_orbitals_cartoon} (indexed with 0 and 1) and the second line uses atomic orbitals L and R through decomposition~\eqref{eq:orbital_coefficients} with optimal coefficients~\eqref{eq:h2_optimized_orbitals_cartoon}.
The representation with atomic orbitals clearly shows, that the electrons in the ground state of the dissociated molecule are spatially separated with one being located at the left and the other at the right atomic orbital. In other words, the two atomic orbitals are never doubly occupied. On the other hand, the representation with molecular orbitals has only double occupied orbitals.
This illustrates how two spatially separated electrons can be represented by using only doubly occupied orbitals in the wavefunction, given that the right choice of orbitals is used.\\

In a larger basis, e.g. H$_2$/6-31G(2,8),  the circuit in Eq.~\eqref{eq:h2_spa_circuit} is extended to
\begin{align}
    \raisebox{1.75cm}{$U_2^8 =$}\includegraphics[width=0.4\textwidth]{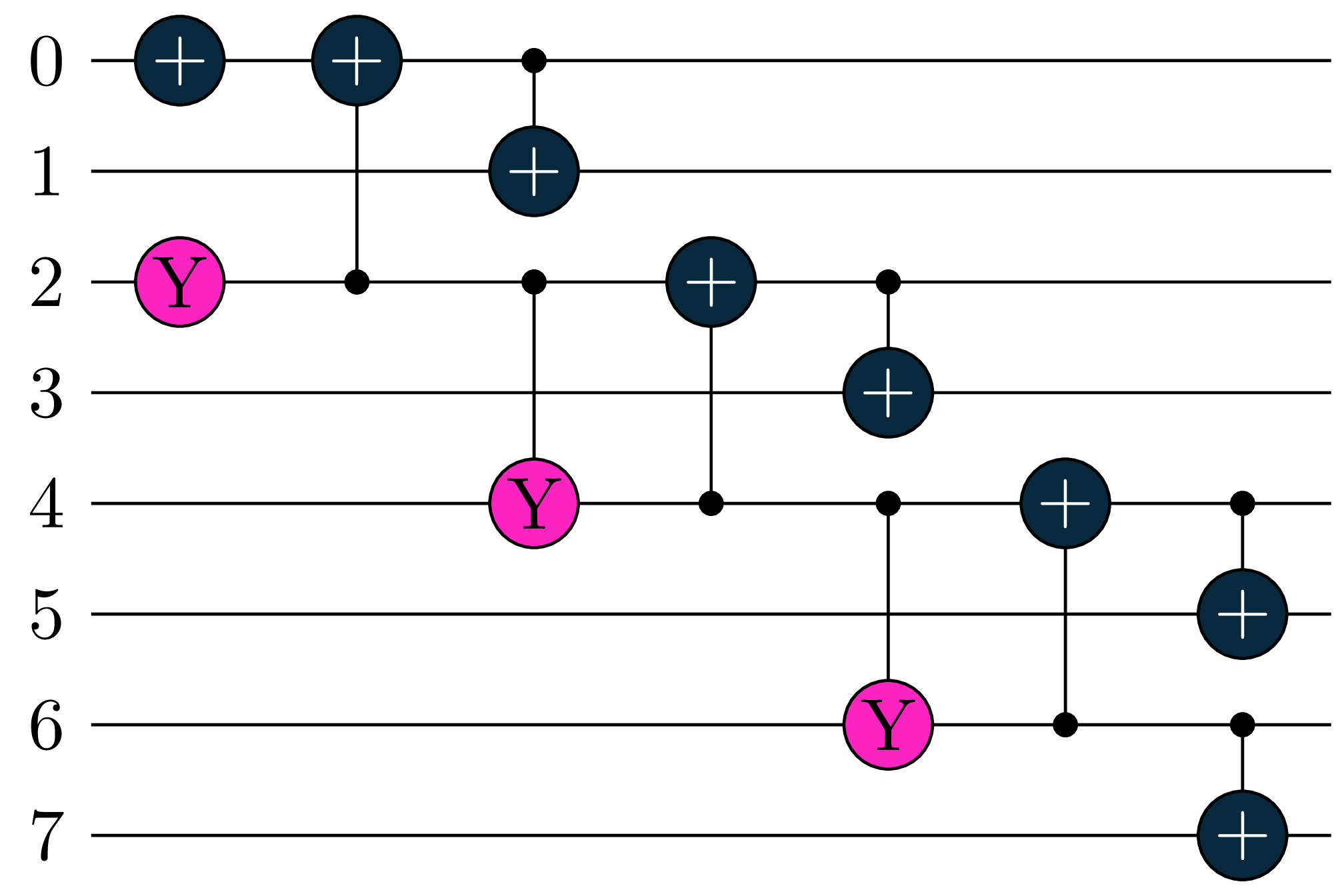}\label{eq:h2_spa_circuit}\;\;.
\end{align}
In general, the $N$ qubit form of this circuit prepares
\begin{align}
    \ket{\psi_2^N} &= U_2^N\ket{0}^{\otimes N} \label{eq:pair_wavefunction}\\ 
    &=\sum_{k=0}^{\frac{N}{2}-1} \cos\lr{\frac{\theta_k}{2}}\prod_{l=0}^{k-1}\lr{\sin\lr{\frac{\theta_l}{2}}}\ket{2^{2k+1}+2^{2k}}\nonumber
\end{align}
with $\ket{2^{k+1}+2^{k}}$ denoting the computational basis states with two 1s at qubits $2k$ and $2k+1$ in binary (e.g. for $k=2$ we have $\ket{48}\equiv\ket{\dots00110000}$). As before, this wavefunction is restricted to doubly occupied orbitals and can describe the molecule exactly given that the orbitals are optimized accordingly. In this case however, the optimized orbitals are not identical to the Hartree-Fock orbitals witnessed by the energy difference in Fig.~\ref{fig:h2_lih_results}.\\

Other than H$_2$ the Helium atom is another obvious two-electron system, here the ground states of the 4-qubit He/6-31G(2,4) or ten qubit He/cc-pVDZ(2,10) can be represented exactly with the same type of circuits as used for H$_2$.
{Meaning the circuit~\eqref{eq:h2_spa_circuit_4} (adapted to the corresponding qubit number) preparing wavefunction~\eqref{eq:pair_wavefunction}. We can essentially view an Helium atom as H$_2$ with zero bond distance (omitting of course the constant nuclear-nuclear potential)}.\\

The beryllium atom and the lithium hydride (LiH) molecule are formally four electron systems but they behave like effective two electron systems where one electron pair remains close to the beryllium, respectively lithium, atom. These so called \textit{core} electrons do usually show only negligible correlation within most basis sets -- the frozen core error within the STO-3G basis is for example below the millihartree threshold for all bond distances (see Fig.~\ref{fig:h2_lih_results} and Fig.~\ref{fig:lih_frozen_core_errors} in the appendix). There are specialized basis sets (\textit{e.g.} the cc-pCVXZ family where the capital C denotes core-polarization) that allow to represent these correlation effects. In the case of LiH there is however still no noteworthy correlation between the pairs, so a separable pair approximation~\cite{kottmann2022optimized} $U_\text{SPA}^{(2,2N)} = U_2^N \otimes U_2^N$ with the individual $U_2^4$ similar to~\eqref{eq:h2_spa_circuit_4} is sufficient for an accurate description (see Fig.~\ref{fig:h2_lih_results} and additional details in the appendix~\ref{sec:lih_details}). In a similar manner we can describe other \textit{single bond} effects, like the dissociation of the C-C bond in C$_2$H$_6$ -- see Ref.~\cite{kottmann2022optimized} where this is explicitly discussed.\\

\subsection{Chemical Graphs and Separated Electron Pairs}
In the previous section it was demonstrated how (near-) exact wavefunctions of (effective) two-electron systems can be prepared with shallow parametrized circuits combined with an optimized orbital representation. In~\eqref{eq:h2_graph} the \textit{chemical graph} (or \textit{Lewis formula}) of the Hydrogen molecule -- where the vertices represent the two hydrogen atoms and the edge represents a \textit{chemical bond} -- was shortly introduced. Graph based representations were applied successfully in quantum optimization approaches in the past~\cite{krenn2021conceptual, kottmann2021quantum, krenn2017quantum} with the graph representations of optical systems specifically developed for this purpose. For chemical systems a graph representation comes more naturally and can be constructed by determining and connecting the \textit{valence} electrons $N^A_e$ for each atom $A$.
In the following a simple scheme that achieves this for neutral molecules with even number of electrons built from atoms of the the first two main-row elements of the periodic table (atomic number $n_A<18$) is briefly described without going into too much detail. This is sufficient for the techniques described in this work but also for the majority of current day research in variational quantum algorithms.\\

Chemical graphs intrinsically use the frozen-core approximation described in the last section and connect only the valence electrons (all electrons that are not core electrons).
Using the periodic system of elements, the number of core electrons for atom $A$ is $N_\text{core}^A = n_B$ with $B$ being the next noble gas with atomic number $n_B \leq n_A$. For the first three rows we have for example
\begin{equation}
    N^A_\text{core} = 
    \begin{dcases}
    0,\; n_A < 2\\
    2,\; n_A < 10\\
    10,\; n_A < 18\\
    \end{dcases},
\end{equation}
with the noble gases Helium ($n_\text{He}=2$) and Neon ($n_{Ne}=10)$.
The number of valence electrons is then
\begin{align}
    N_\text{valence}^A = n^A - N_\text{core}^A. 
\end{align}
Now each atom in the molecule is assigned $N_\text{valence}^A$ connectors and all that is left is to connect them to obtain the edges of the graph. Note, that the connectors can also form loops (so called \textit{lone pairs}) which is often a preferred choice once $N_\text{valence}^A>4$.~\footnote{In particular this means: $H,He,Li,B,C$ have no lone-pairs, $N$ has one, $O$ has two, $F$ has three, and Ne has four lone-pairs that are, in this work, considered part of the frozen core. In order to keep it simple atoms with lone pairs are mostly avoided in the examples of this work. The atomic graph of BeH$_2$ can however be seen as having a lone-pair on the Be atom.} Usually the most important chemical graph is the one where all atoms are maximally connected to their closest neighbors (see the linear H$_4$ example introduced further below). Note however, that the most important graph is not always unique (a famous example being conjugated ring systems~\cite{kottmann2023compact} like H$_4$ or Benzene).\\

Once a chemical graph is determined we can construct a circuit from it by treating all edges as separable electron pairs whose wavefuncions can be prepared by the same circuits as in~\eqref{eq:h2_spa_circuit} and the optimal orbitals are determined as the linear combination of basis orbitals that yield the lowest energy for this circuit.
This is the so called separable pair approximation (SPA)~\cite{kottmann2022optimized} that was originally introduced in combination with system-adapted orbitals determined through a surrogate model~\cite{kottmann2020reducing, kottmann2020direct} and is related to classical (generalized) valence-bond approaches (see for example~\cite{shaik2022nature, goddard1973generalized, vbtheory} or ~\cite{larsson2020minimal} for a modern approach in the context of tensor networks).\\
In Heuristic~\ref{heuristic:SPA} an SPA is constructed from a chemical graph of a molecule (note that the first $
\textsc{cr}_y$ can be implemented as just $\textsc{r}_y$ -- see~\ref{eq:h2_spa_circuit}). Here the pair ranks $R_\text{e}$ determine how many spatial orbitals are assigned to each edge -- the H2/STO-3G(2,4) system with circuit~\ref{eq:h2_spa_circuit_4} for example has a pair rank of 2 and the H2/6-31G(2,8) with circuit~\ref{eq:h2_spa_circuit} has a pair rank of 4. A minimally correlated approach (employed within most examples in this work) is to set $R_\text{e}=2$ for all edges. Like most variational quantum approaches the separable pair approximation described in Heuristic~\ref{heuristic:SPA} is not an exact algorithm as successful convergence to the best solution for a given graph depends on the initial guess for the optimal orbitals. In this work we explicitly exploit insight from the molecular graph in order to generate good initial guesses as demonstrated within the examples below. 
{In particular the instruction \textit{assign orbitals to edge} in Heuristic~\ref{heuristic:SPA} can be realized with different strategies. In this work, we are assigning atomic orbitals to the corresponding vertices $v$. In the case of hydrogenic systems, this is straightforward, as there is just a single vertex for each atom, meaning that for every edge $e=(v_k,v_l)$ we assign the atomic orbitals centered at the atoms corresponding to the two vertices that form the edge. For other systems, such as BeH$_2$ we need to decide which atomic orbital to map to which vertex, as for example the Be atom will result in two vertices. This will be illustrated explicitly in the next section.}\\

As it was shown in~\cite{kottmann2022optimized} the SPA wavefunctions are classically tractable with linear memory requirement with respect to the number of qubits. This can be seen from Eq.~\eqref{eq:pair_wavefunction} that describes a single electron pair in $R_\text{e}$ spatial orbitals with $R_\text{e}$ coefficients and Heuristic~\ref{heuristic:SPA} that constructs an SPA wavefunction through $|G|$ electron pairs, with $|G|$ denoting the number of edges in the graph. The total wavefunction can therefore be represented by at most $\sum_{\text{e}}R_\text{e} \leq \max\lr{R_\text{e}} |G|$ real coefficients -- for a minimal correlated orbital set that would for example be $2|G|$. If second order orbital optimization is employed the overall procedure stays classically tractable, with the orbital optimization scaling at most quartic with system size.~\cite{mizukami2020, sokolov2020quantum}
The numerical results of Ref.~\cite{kottmann2022optimized} indicate that Graph-SPAs are a good approximation of the wavefunction if the chemical graph is dominant in its importance. The importance of chemical graphs usually cannot be quantified explicitly, but they can be ordered in their relative importance based on chemical intuition and arguments from Lewis theory (see Sec.~\ref{sec:examples} for explicit examples). It might be argued, that in fact the SPA wavefunction quantifies the graph (e.g. through its fidelity with respect to the exact ground state, or via the energy expectation value) and not the other way around.

\begin{heuristic}
\caption{Graph-SPA}\label{heuristic:SPA}
\begin{algorithmic}
\Require edges $E$, pair ranks $R_{e}$, orbitals  $\left\{\phi_k\right\}$
\State initialize circuit $U \gets \left\{\right\}$
\For{$e \in E$}:
\State assign orbitals to edge: $e 
\rightarrow \left\{\phi_k\right\}$
\State assign qubits $\left\{q_{2k}, q_{2k+1}\right\} \gets \left\{\phi_{k\uparrow}, \phi_{k\downarrow}\right\} $
\State $U 
\gets U \cup  \textsc{not}(q_0)$
\For{$0\leq k\leq R_{e}-2$}:
\State $U \gets U \cup \textsc{cr}_y(q_{2k},q_{2k+2},\theta_{q_{2k}})$ 
\State $U \gets U \cup \textsc{cnot}(q_{2k+2},q_{2k})$
\State $U \gets U \cup \textsc{cnot}(q_{2k},q_{2k+1})$
\EndFor
\State $U \gets U \cup \textsc{cnot}(q_{2(R_{e}-1)},q_{2(R_{e}-1)+1})$
\EndFor
\State construct initial guess for optimal orbitals
\State initialize Hamiltonian H
\While{energy is changing}
\State $\boldsymbol{\theta^*} \gets \argmin_{\boldsymbol{\theta}} \langle H \rangle_{U(\boldsymbol{\theta})}$
\State optimize orbitals
\State update Hamiltonian
\EndWhile
\State \Return $U$, $H$, $\boldsymbol{\theta}^*$
\end{algorithmic}
\end{heuristic}

\subsection{Quantum Circuits from multiple Graphs}
As SPAs are classically tractable, it is clear that they cannot represent all molecules with arbitrary precision. Obvious candidates are molecules with multiple chemical graphs necessary for a satisfactory description. In Sec.~\ref{sec:examples} explicit examples for such systems are given.
One way to incorporate more than one chemical graph in the design of a quantum circuit is by rotating the orbitals that represent one graph into orbitals that represent another graph, apply electron correlators, and rotate back into the original frame. 
In a way this can be seen as an autoencoder~\cite{romero2017quantum} with the orbital rotators as encoder and the corresponding adjoint as decoder. The main difference in spirit to~\cite{romero2017quantum} is that the goal here is not compression and that there is no ``training'' involved.\\
The procedure is formalized in Heuristic~\ref{heuristic:multigraph} and its success depends on the right choice and initialization of orbital rotators, which will again be tackled through insight from the molecular graph.\\ Although there is also a heuristic choice in electron correlators, in practice it often suffices to use electron pair-excitations (the simplest electronic correlators) with angles initialized to zero. In the next section we will walk through some explicit examples how Heuristic~\ref{heuristic:multigraph} can be applied in practice and how successful circuits can be transferred between different molecules with similar graph representation.
\begin{heuristic}
\caption{Multi-Graph Circuit}\label{heuristic:multigraph}
\begin{algorithmic}
\Require: Chemical graphs $\mathcal{G}$
\State Select $\tilde{G}$=$(V,E) \in \mathcal{G}$
\State Construct $U_\text{SPA}$ and $H$ according to Heuristic~\ref{heuristic:SPA}.
\State $U \gets U_\text{SPA}$
\For{$G \in \mathcal{G}/\tilde{G}$}: 
    \State Construct orbital rotations $U_G$
    \State Choose correlator block $U_C$
    \State $U \cup U_G U_CU_G^\dagger$
    \State $\boldsymbol{\theta^*} \gets \argmin_{\boldsymbol{\theta}}\langle H \rangle_{U(\boldsymbol{\theta})}$
\EndFor
\State \Return $U, \boldsymbol{\theta}^*$
\end{algorithmic}
\end{heuristic}

\section{Examples}\label{sec:examples}
In the following we will apply the concepts discussed in the previous section to explicit examples beyond (effective) two-electron systems.
This will demonstrate explicitly how suitable orbital rotations and useful initial guesses can be constructed by simple intuition drawn from the chemical graphs of the molecules. We start with small strongly correlated multi-electron systems H$_4$ and H$_6$ that are challenging for classical methods and finally show how the circuits constructed for H$_4$ can be transferred to other effective four electron systems like BeH$_2$ and the $\pi$-system of C$_4$H$_6$.

\subsection*{Linear H$_4$}
A simple example for a electronically non-trivial system is the linear H$_4$ where four hydrogen atoms are placed on a straight line with 1.5~{\AA}~ inter-atomic distance
\begin{align}
    \includegraphics[width=0.3\textwidth]{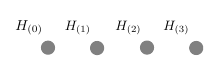}.
\end{align}

The most intuitive chemical graph is to view the system as two hydrogen molecules
\begin{align}
    \includegraphics[width=0.3\textwidth]{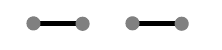}\label{eq:h4_graph_two_bonds}.
\end{align}
We will now represent this chemical graph with a quantum circuit in a minimal orbital basis, \textit{i.e.} H$_4$/STO-3G(4,8), obtained by placing a single s-type orbital on each hydrogen atom, in the same way as in~\eqref{eq:h2_atomics_cartoon}. 
A quantum circuit that represents this chemical graph on eight qubits is build from two separated circuits that represent the individual H$_2$ bonds
\begin{align}
    \raisebox{1.5cm}{$U_\text{SPA}^{(4,8)} = U_2^4 \otimes U_2^4 =$} \includegraphics[height=0.15\textheight]{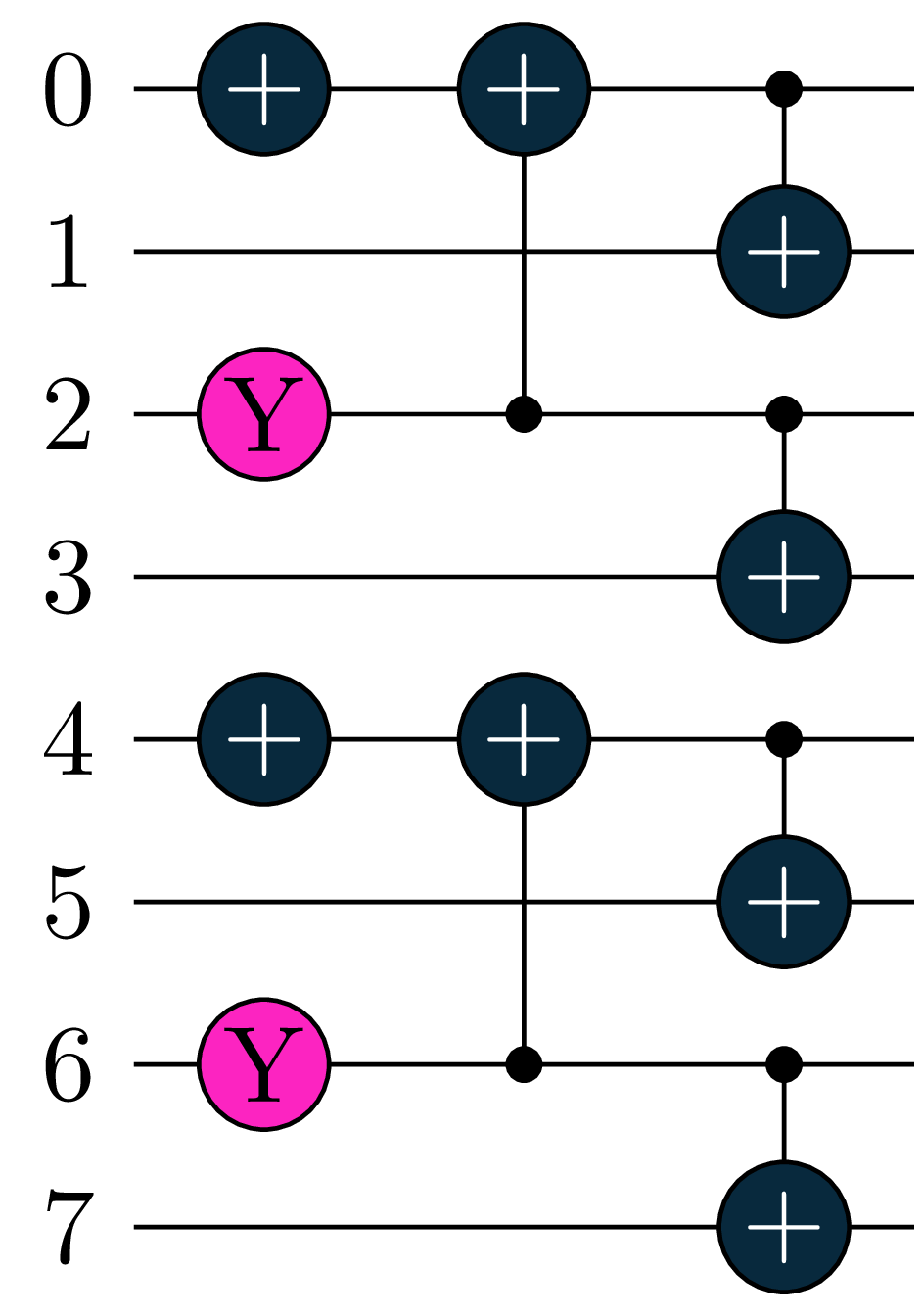}\label{eq:h4_spa}
\end{align}
where the individual $U_2^4$ circuit are identical to~\eqref{eq:h2_spa_circuit_4} that described an individual hydrogen molecule.
As before, the atomic basis functions need to be rotated to optimal linear combinations for the wavefunction created by the $U_\text{SPA}^{(4,8)}$ circuit.  An intuitive initial guess is to rotate them to localized orbitals that represent individual hydrogen molecules as in~\eqref{eq:h2_optimized_orbitals_cartoon}. Starting from (orthonormalized) atomic orbitals, this rotation acts as
\begin{align}
    \includegraphics[width=0.2\textwidth]{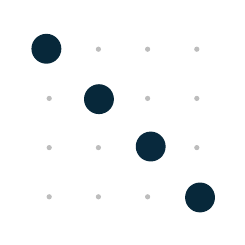}\raisebox{1.5cm}{$\xrightarrow{U_{\text{R}_0}}$}
    \includegraphics[width=0.2\textwidth]{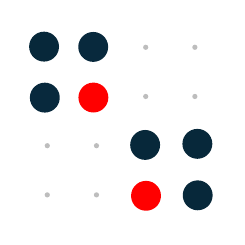}.
\end{align}
With this initial guess for the orbitals, the wavefunction prepared by $U_\text{SPA}^{(4,8)}$ with optimized angle has an error of 40 millihartree in energy with respect to exact diagonalization. If we further optimize the orbitals through standard procedures the error can be reduced to 16 millihartree (denoted  as SPA in Tab.~\ref{tab:h4_results}) with optimal orbitals that look like
\begin{align}
\includegraphics[width=0.2\textwidth]{h4_orbitals_two_bonds.pdf} \raisebox{1.5cm}{$\xrightarrow[\text{optimization}]{\text{orbital}}$}
\includegraphics[width=0.2\textwidth]{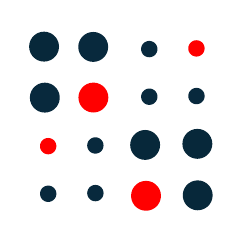}\label{eq:h4_optimized_orbitals_cartoon}
\end{align}
(see the appendix for the explicit coefficients).
As before we can generate a transformed Hamiltonian with the optimized orbitals instead of including the rotation in the circuit {(see appendix~\ref{sec:orbital_rotation_circuits} and Sec.~C.a of Ref.~\cite{kottmann2022optimized} for details on the implementation)}, so in the following the orbitals in \eqref{eq:h4_optimized_orbitals_cartoon} will be our basis.\\
Due to the equidistant placing of the hydrogen atoms, the chemical graph~\eqref{eq:h4_graph_two_bonds} alone does not describe the situation accurately. Another possibility would be to have a central hydrogen molecule surrounded by two hydrogen atoms
\begin{align}
    \includegraphics[width=0.3\textwidth]{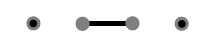}\label{eq:h4_chemical_graph_central_bond},
\end{align}
where we could also connect the outer atoms to form a spatially separated singlet pair similar to the solution of the stretched H$_2$ discussed in the previous section. We will chose not to do so in order to form a shallower and more localized circuit.
The strategy is now, to rotate the optimized orbitals in~\eqref{eq:h4_optimized_orbitals_cartoon} to a suitable representation for the chemical graph in~\eqref{eq:h4_chemical_graph_central_bond} with the corresponding unitary $U_\text{R}$, then correlate the orbitals that describe the central bond (between orbitals 1 and 2) with a unitary $U_\text{C}=U_{\text{C}_{1}^{2}}$, and finally rotate back to the original frame with $U_\text{R}^\dagger$. The full circuit, simply abbreviated as SPA+ is then
\begin{align}
    U_\text{SPA+} =  U_\text{R}^\dagger U_{\text{C}_{1}^{2}} U_\text{R}U_\text{SPA}^{(4,8)}\label{eq:def_spa+}
\end{align}
which is the same circuit as shown in Fig.~\ref{fig:overview_cartoon}.
We use a pair-restricted double excitation that transfers spin-paired electrons between two spatial orbitals (e.g. from orbital 1 to orbital 2) denoted as
\begin{align} \raisebox{0.85cm}{$U_{\text{C}_{1}^{2}}=$} \includegraphics[height=0.09\textheight]{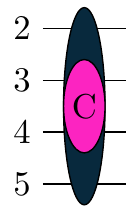} \raisebox{0.75cm}{$=$} \includegraphics[height=0.09\textheight]{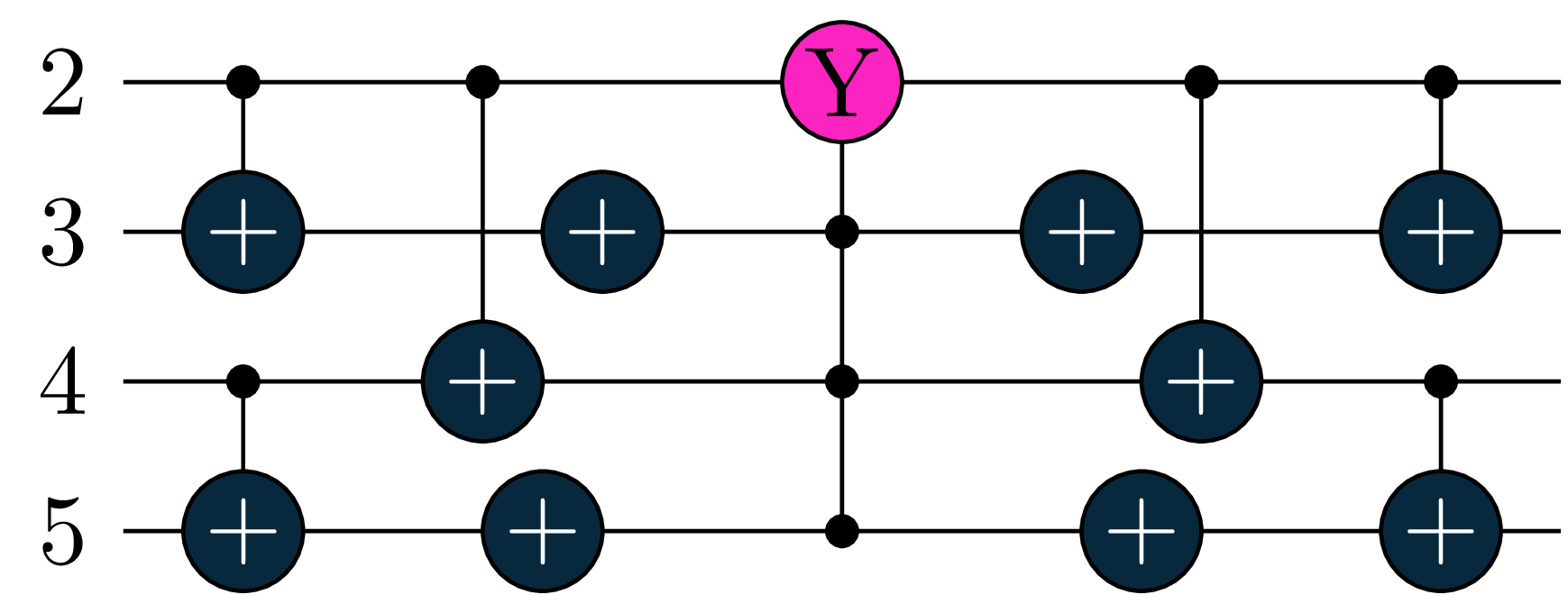}\label{eq:h4_u1c},
\end{align}
where the circuit represents an optimized form of the Jordan-Wigner encoded unitary $e^{-i\frac{\theta}{2} G}$ with the generator $G=i\left(a^\dagger_2 a_3 a^\dagger_4 a_5 - h.c.\right)$. Note, that the correlator type is independent of the chemical graph. In this case the pair-restricted double excitations were chosen since they are the cheapest fermionic correlators that have been applied in several other works.~\cite{yordanov2020efficient, kottmann2022optimized, anselmetti2021local, lee2018generalized}
In order to get a good initial guess for the parameters of the rotation circuit $U_\text{R}$ we split its construction into two parts: First (approximately) rotating back to the local frame $U_{\text{R}_0}^\dagger$ and second forming the desired orbitals with an initial guess formed from the atomic orbitals through the unitary $U_{\text{R}_1}$. This corresponds to the following pattern
\begin{align}
    \includegraphics[width=0.2\textwidth]{h4_two_bonds_optimized.pdf}\raisebox{1.5cm}{$\xrightarrow{U^\dagger_{\text{R}_0}}$}
    \includegraphics[width=0.2\textwidth]{h4_orbitals_atomic.pdf}\raisebox{1.5cm}{$\xrightarrow{U_{\text{R}_1}}$}
    \includegraphics[width=0.2\textwidth]{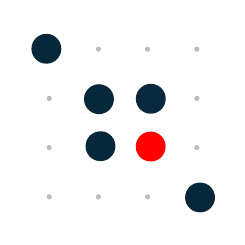},\label{eq:h4_rotator_sketch}
\end{align}
using $U_\text{R$_1$} = U_{\text{R}_{1}^{2}}$ and $U_{\text{R}_0} = R_{\text{R}_{0}^{1}} U_{\text{R}_{2}^{3}}$ with right-hand-sides defined as in Eq.~\eqref{eq:h2_R}.\\
With this strategy we observe rapid convergence and further reduction in the energy error to 8 millihartree displayed in Tab.~\ref{tab:h4_results} along with a small comparison to the prominent $k$-UpCCGSD~\cite{lee2018generalized} approach that uses the same building blocks. Here, we used fixed initial angles (set to zero) for the $k$-UpCCGSD optimization in order to have a comparable and reproducible setting. The SPA and SPA+ circuits outperform the $k$-UpCCGSD flavors in all metrics (energy, fidelity, number of parameters, number of cnots, circuit depth, runtime -- indicated with iteration counts). Another interesting effect can be observed in the fidelities with the H$_4$ ground state which is better for UpCCD compared to UpCC(G)SD without being reflected in the corresponding energies. The reason for this being overlaps with high energy excited states in the UpCCD wavefunction that increase the expected value of the energy. The fidelity drop from UpCCD to UpCC(G)SD witnesses the increased complexity of the optimization landscape that make the H$_4$ system hard to converge without exploiting its chemical structure. Note that UpCCGSD and 2-UpCCGSD can achieve higher accuracies in energy and fidelity for better initial angles. Repeated optimization runs with randomized initial angles can increase the accuracy, UpCCGSD is however not able to come close to the classically tracable SPA while 2-UpCCGSD can achive better energies and fidelities as SPA+ with significantly increased computational cost and an unreliable procedure. ADAPT-VQE with an operator pool build from the UpCCGSD operations achieves the same accuracy as 2-UpCCGSD with a more efficient circuit at the price of more iterations in the parameter optimization and 12 operator screening rounds. Despite having access to the same set of operations as SPA and SPA+, ADAPT(UpCCGSD) does not end up with a comparable circuit with respect to all metrics. An intuitive explanation is, that the adaptive procedure cannot locally detect the $U_\text{R}U_\text{C}U_\text{R}^\dagger$ motif as, for example, $U_\text{R}$ alone will not improve the energy, let alone parts of $U_\text{R}$. In the same manner, $U_\text{C}$ has not the same effect without the basis change before and after.

\subsection*{Linear H$_6$}
A similar system to the linear H$_4$ is the linear H$_6$ with the same inter-atomic distances of 1.5\AA. We will now construct circuits that prepare good approximations for the groundstate of H$_6$/STO-3G(6,12) by applying the same strategy as for the H$_4$/STO-3G(4,8). The two main chemical graphs are in this case
\begin{align}
    \includegraphics[width=0.3\textwidth]{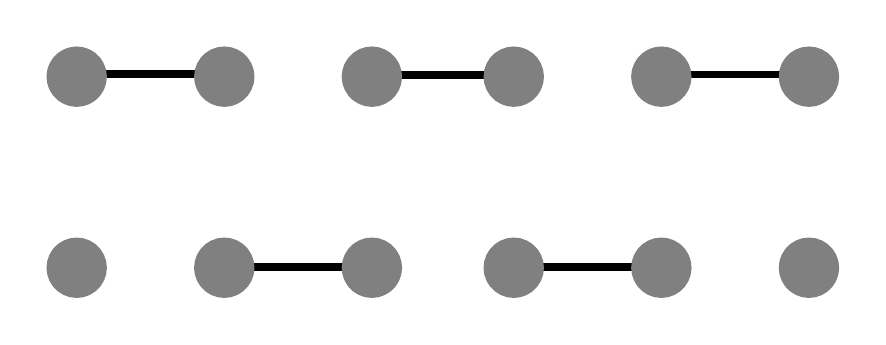}\label{eq:h6_configurations}
\end{align}
and the corresponding SPA+ circuit is constructed as $U_6^{12} = U_2^4\lr{\theta_a} \otimes U_2^4\lr{\theta_b} \otimes U_2^4\lr{\theta_c}$ with orbitals that resembles the first graph in~\eqref{eq:h6_configurations}. As before further simplifications based on symmetry $(\theta_a = \theta_c)$ could be made but will not be included in the analysis of the associated computational cost.\\
In order to represent the second graph in~\eqref{eq:h6_configurations} we follow the same strategy as before by constructing a orbital rotation unitary $U_\text{R}$ in two parts: One rotating to a localized frame followed by a second one that forms the orbitals which represent the central H-H bond. The rotation is again followed by a correlating unitary $U_\text{C}$ given by a pair-excitation on the central H-H followed by a rotation to the original frame. The so constructed SPA+ circuit is shown in full in~\eqref{eq:h6_spa+} in the appendix and the obtained results are given in Tab.~\ref{tab:h6_results}. The observed behaviour is analog to the linear H$_4$.

\begin{table*}
    \centering
    \begin{tabular}{lrcrrrr}
\toprule
method     & error &$F$ & $N_\text{v}$ & cnots & depth & iter \\
\midrule
SPA       & 16 & 94 &  2 &   6 &   3 &   17$^*$ \\
SPA+       & 8 &  96 &  6 & 116 & 131 &  10 \\
SPA+X  & 0 & 100 & 19 & 294 & 317 & 160 \\
\midrule
UpCCD      & 103 & 88 & 4 &  20 &  26 &   8 \\
UpCCSD     & 86 & 74 &12 & 148 & 193 &  13 \\
UpCCGSD    & 86 & 74 &18 & 188 & 254 &  13 \\ 
2-UpCCGSD  & 32 & 90 &36 & 432 & 540 &  48 \\
\midrule
ADAPT(UpCCGSD) & 32 & 90 & 12 & 448 & 442 & 113 \\
ADAPT(UCCGSD) & 0 & 100 & 21 & 1360 & 1705 & 58 \\
\midrule
\multicolumn{1}{l}{random start}    & \multicolumn{2}{c}{$\langle \text{error}\rangle$ (best)} & \multicolumn{2}{c}{$\langle F \rangle$ (best)} & \multicolumn{2}{c}{ $\langle \text{iter}\rangle$ }\\ 
\midrule
\multicolumn{1}{l}{UpCCGSD}    & \multicolumn{2}{c}{37(16)} & \multicolumn{2}{l}{79(95)} &  \multicolumn{2}{c}{\phantom{0}57} \\ 
\multicolumn{1}{l}{2-UpCCGSD}    & \multicolumn{2}{c}{4(2)} & \multicolumn{2}{l}{98(99)} &  \multicolumn{2}{c}{127} \\ 
\bottomrule
\multicolumn{6}{l}{$^*$\footnotesize includes orbital optimization}
    \end{tabular}
    \caption{Performance of molecular circuits (SPA, SPA+, and SPA+X with additional rotators respectively correlators -- see Fig.~\ref{fig:systematic_errors_h4_c4h6_beh2}) compared with standard UCC methods for the linear H$_4$ molecule (bond distance 1.5{\AA}). Shown errors denote the difference to FCI/STO-3G in millihartree, iter denotes the total number of BFGS iterations and $N_\text{v}$ the number of variational parameters. Fidelities $F$ are given as percentage with respect to the exact STO-3G ground state. The displayed results with random initial values were collected over 10 runs with $\langle \cdot \rangle$ denoting averages, additional the best results are given in parenthesis. SPA(+X) and (2-)UpCC(GS)D circuits are compiled with techniques of~\cite{kottmann2022optimized} while ADAPT uses standard compilation. The ADAPT(UpCCGSD) circuits in particular are therefore expected to have lower cnot counts and depths as similar optimization can be applied to parts of the circuit. Note that circuit-depths can be further reduced through more efficient compilation of the orbital rotations~\cite{kivlichan2018quantum, google2020hartree}}
    \label{tab:h4_results}
\vspace{0.5cm}
    \begin{tabular}{lrrrrr}
\toprule
method     & error & $N_\text{v}$ & cnots & depth & iter \\
\midrule
SPA        & 33 &  3 &   9 &   3 &   4 \\
SPA+       & 18 & 10 & 197 & 131 &  8 \\
SPA+X      & 5 & 32 & 489 & 317 & 95 \\
\midrule
UpCCD      & 188 &  9 &  42 &  42 &   8 \\
UpCCSD     & 184 & 27 & 474 & 561 &  12 \\
UpCCGSD    & 141 & 45 & 626 & 760 &  39 \\
2-UpCCGSD  & 133 & 90 & 1396 & 1536 &  51 \\
\midrule
Adapt(UpCCGSD) & 59 & 65 & 2352 &  2119 & 3391 \\
\bottomrule
    \end{tabular}
    \caption{Performance of chemically targeted circuits (SPA, SPA+, and SPA+X) compared with standard UCC methods for the linear H$_6$ molecule. SPA+ refers to the circuit in~\eqref{eq:h6_spa+} and SPA+X to a similar circuit as for H$_4$ described in Tab.~\ref{tab:h4_results}. Shown errors denote the difference to FCI/STO-3G in millihartree, \textit{iter} denotes the total number of BFGS iterations and $N_\text{v}$ the number of variables in the circuit. See Tab.~\ref{tab:h4_results} for comments on cnot counts.}
    \label{tab:h6_results}
\end{table*}

\subsection*{From H$_4$ to BeH$_2$ and C$_4$H$_6$}
Other than the rather artificial H$_4$ toy model, the BeH$_2$ molecule and the $\pi$-system of C$_4$H$_6$ come closer to real-life examples. In the following we will see how they can be treated analog to the H$_4$ system by using the same circuits and intuition. We start by looking at the chemical graphs that represent the BeH$_2$ system
\begin{align}
    \includegraphics[width=0.25\textwidth]{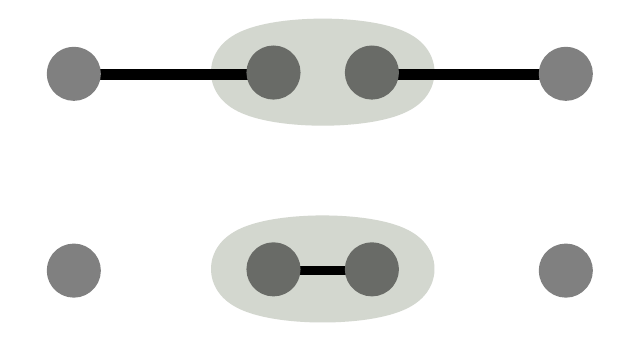}\label{eq:beh2_graphs}
\end{align}
where we have depicted the central beryllium atom with two vertices, representing the two valence electrons.
The first graph in~\eqref{eq:beh2_graphs} represents the molecular system with two Be-H bonds and the second the isolated atomic systems; their relative importance will vary with the Be-H inter-atomic distances. Note that the graphs are identical to the H$_4$ graphs in~\eqref{eq:h4_chemical_graph_central_bond} and ~\eqref{eq:h4_graph_two_bonds}.
Using the frozen-core approximation and a minimally correlated orbital basis -- meaning that we use the same number of spatial orbitals as we have active (i.e. non-frozen) electrons, we arrive at an eight qubit representation of BeH$_2$/STO-3G(4,8). In this case, we can remove the $p_y$ and $p_x$ orbitals form the STO-3G set -- assuming the Be-H bonds are aligned in $z$ direction), as they will not contribute to the optimized orbitals due to symmetry reasons. As for H$_4$ the remaining orbitals are then optimized with respect to the $U_{\text{SPA}}^{(4,8)}$ wavefunction. A suitable guess for the molecular and atomic case is
\begin{align}
    \includegraphics[height=0.15\textheight]{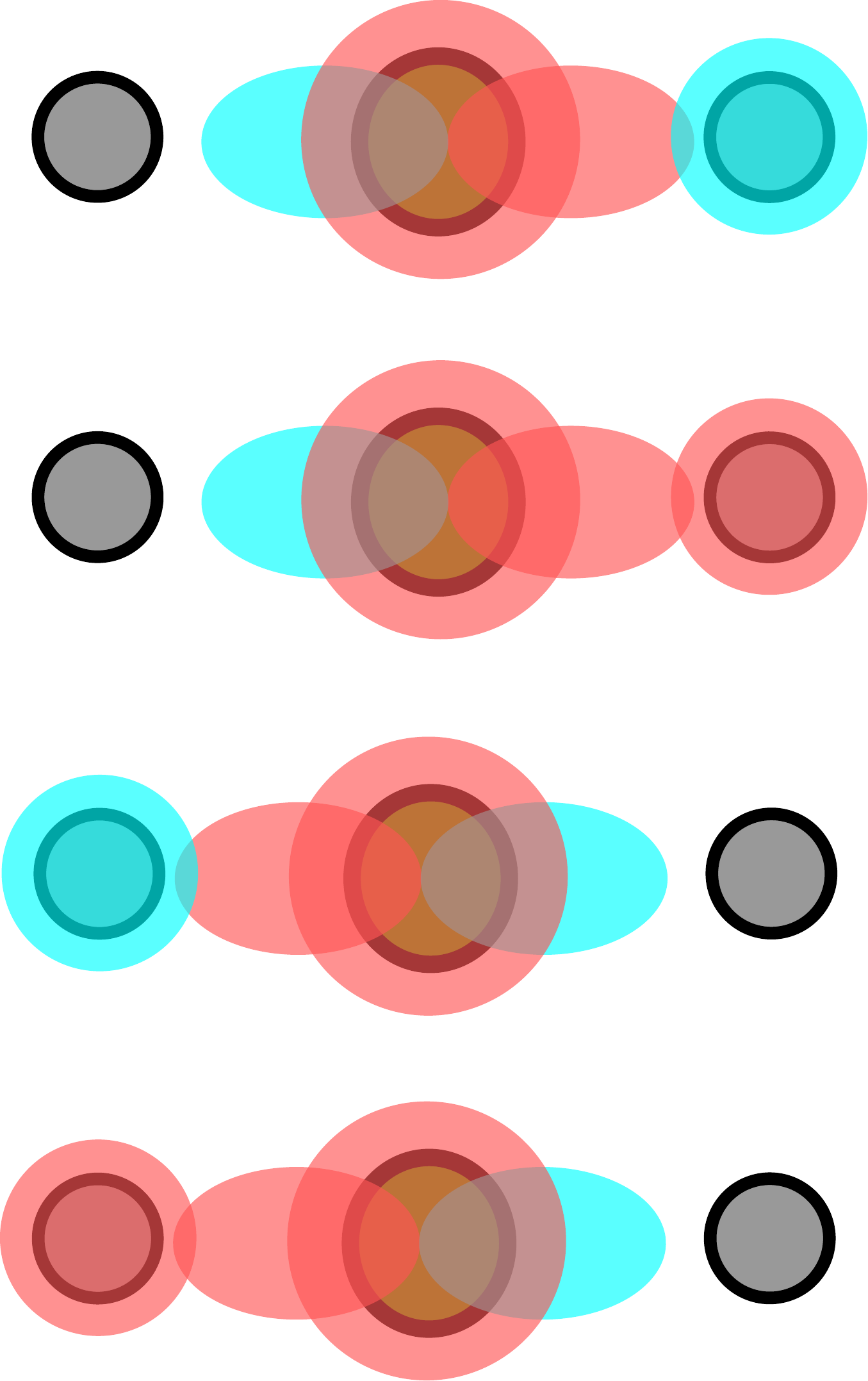},\quad\quad\quad\quad\quad\quad
    \includegraphics[height=0.15\textheight]{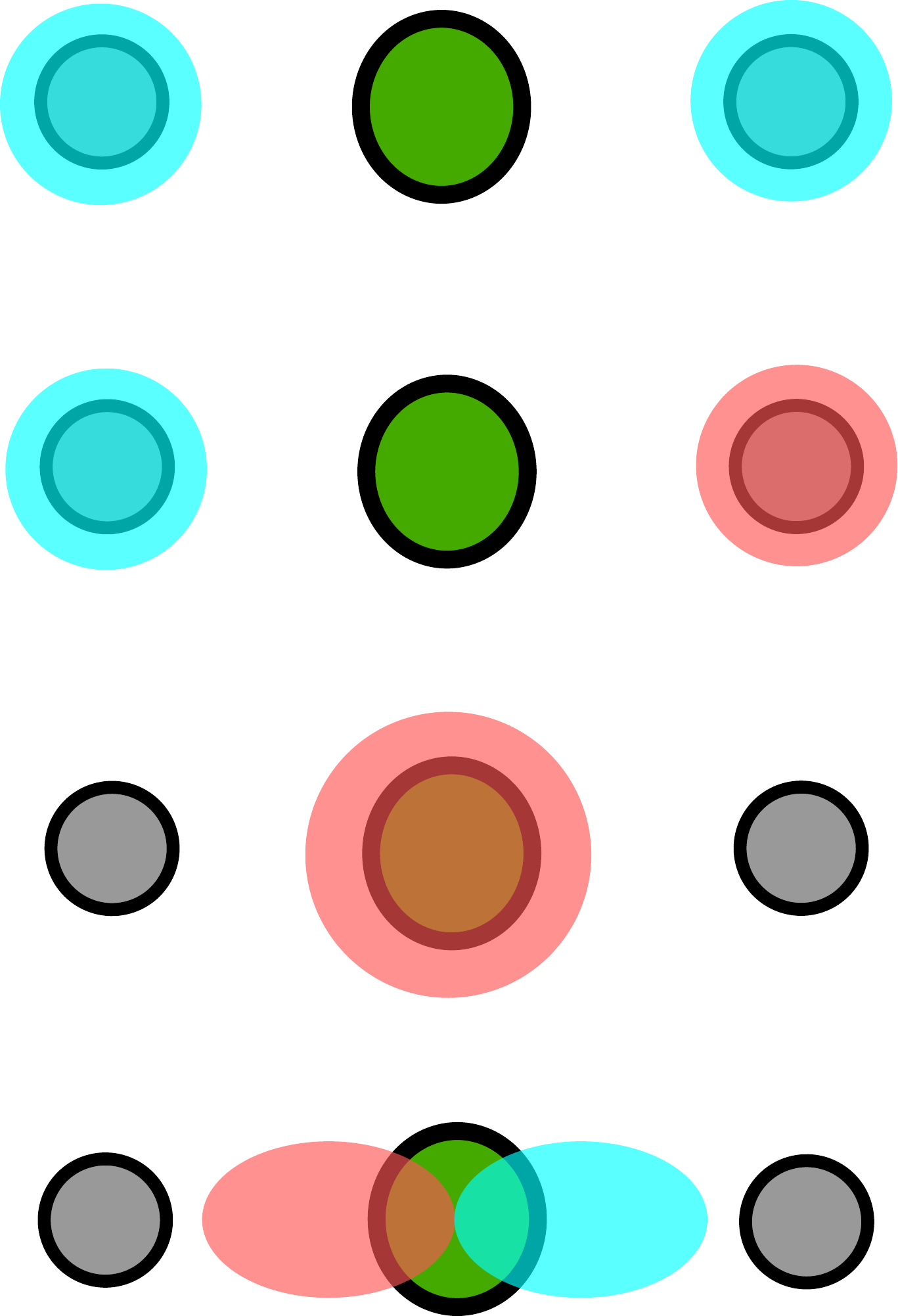}\label{eq:orbitals_beh2}
\end{align}
where $s$ and $p_z$ orbitals are depicted in equal weighted superpositions with the atomic case on the right and the molecular case on the left hand side of~\eqref{eq:orbitals_beh2}. The beryllium atom is  the green circle and the shades of blue and red in the orbitals again denote positive and negative regions of the orbitals.
The optimized coefficients can be found in the appendix.\\

In the same way, we can treat the $\pi$-system of C$_4$H$_6$. This denotes the four electrons in orbitals spanned by the 4 $p_z$ (assuming the molecule lies in the $x,y$ plane) basis function of the STO-3G set. The other 26 electrons doubly occupy their corresponding frozen orbitals determined through Hartree-Fock. We will denote this active space of the so called $\pi$-system as C$_4$H$_6$/STO-3G(4,8) where the optimal $\pi$-orbitals are, again, determined through the $U_{\text{SPA}}^{(4,8)}$ wavefunction and will look similar to the H$_4$ orbitals~\eqref{eq:h4_optimized_orbitals_cartoon} just with $p_z$-orbitals instead of $s$-orbitals, so that the identidcal initial guesses can be used for the two systems. The coefficients of the optimized $\pi$-orbitals can be found in the appendix. Note that those $\pi$-orbitals differ from the Hartree-Fock description of the $\pi$-system.\\

\begin{figure}
\centering
\begin{tabular}{cc}
    \includegraphics[width=0.75\textwidth]{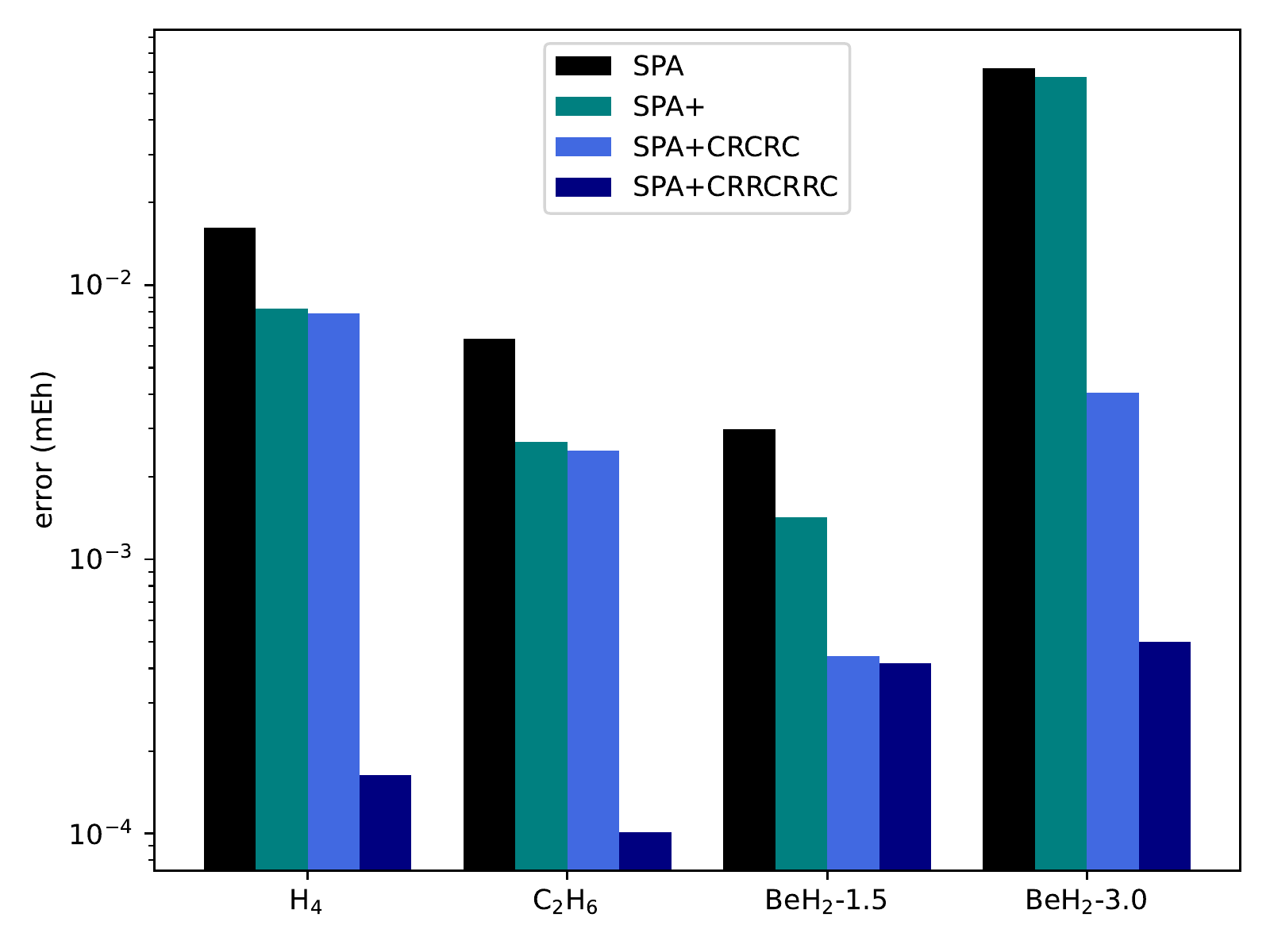}&
\end{tabular}
\caption{Errors for H$_4$/STO-3G(4,8), BeH$_2$/STO-3G(4,8) and the $\pi$ system of C$_4$H$_6$/STO-3G(4,8). The used energies are from quantum circuits that are identical for all three systems. The SPA circuit is given in Eq.~\eqref{eq:h4_spa} and SPA+ denotes the circuit in Fig.~\ref{fig:overview_cartoon}. The other two variants follow the same pattern as in Fig.~\ref{fig:overview_cartoon} but use more general, locally restricted, correlator blocks abbreviated as C, and extended rotation layers abbreviated as RR, both shown in~\eqref{eq:RR-circuit}.} \label{fig:systematic_errors_h4_c4h6_beh2}
\end{figure}

In Fig.~\ref{fig:systematic_errors_h4_c4h6_beh2} we show the performance of identical circuits on H$_4$/STO-3G(4,8), C$_4$H$_6$/STO-3G(4,8) and BeH$_2$/STO-3G(4,8) with equilibrium (1.5{\AA}) and stretched (3.0{\AA}) Be-H bond distances. As circuits we use the SPA circuit (that also determines the optimized orbitals) in~\eqref{eq:h2_spa_circuit_4}, the SPA+ circuit displayed in Fig.~\ref{fig:overview_cartoon} and two extensions where we replace the single central pair-correlator~\eqref{eq:h4_u1c} by a more general block $U_\text{C}^4$ containing four pair correlators with individual parameters which are then placed before and after the rotations. In a second step we also improve the freedom in the orbital rotations by replacing the sequences in Fig.~\ref{fig:overview_cartoon} a more flexible block $U_\text{RR}$,
which we abbreviate in Fig.~\ref{fig:systematic_errors_h4_c4h6_beh2} with RR while using R for the original sequence in Fig.~\ref{fig:overview_cartoon}.
The circuit templates for the modified correlator and rotation blocks look like
\begin{align}
    \raisebox{1.7cm}{$U_\text{C}^4 =$} \includegraphics[height=0.25\textwidth]{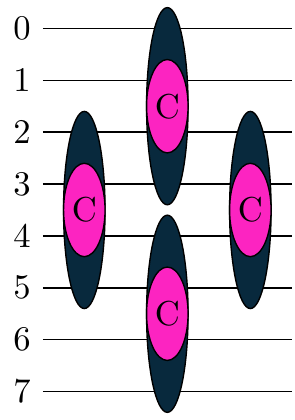},\quad
    \raisebox{1.7cm}{$U_\text{RR} =$}
    \includegraphics[height=0.25\textwidth]{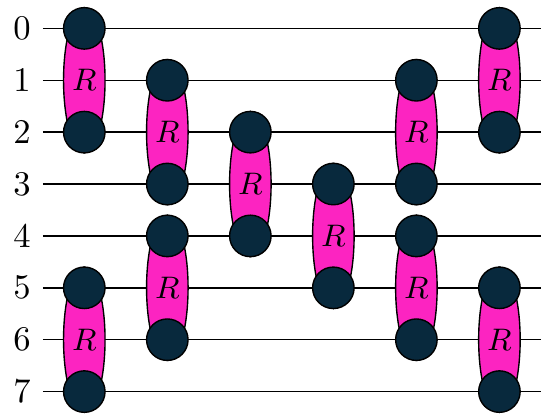}\label{eq:RR-circuit}.
\end{align}
This allows for a systematic improvement of the energy below the millihartree threshold for all four systems.

\section{Limitations}
As Heuristic~\ref{heuristic:multigraph} is relying on chemical insight in order to provide important chemical graphs, it is clear, that the approach will not be able to describe all molecules with sufficient accuracy. Besides from transition metal compounds -- which are fairly unexplored in the context of variational quantum algorithms -- a simple example is the nitrogen molecule N$_2$ where only one reasonable chemical graph can be constructed, but the corresponding Graph-SPA energy still differs significantly from the exact ground state energy even at the equilibrium bond distance. This holds true for a minimal basis (STO-3G) as well as for directly determined MRA-PNOs and can be observed already in the (6,12) active space that freezes out the lone-pairs and includes only the triple bond as it was shown for N$_2$/MRA-PNO(6,12) before~\cite{kottmann2022optimized}. It can be speculated, that the energy can be brought down further with the right intuition about graphs and corresponding orbitals. Indeed a successful trial could be achieved reducing the energy error of N$_2$/STO-3G(6,12) close to equilibrium bond distance from around 40 millihartree (single graph SPA) to 3 millihartree (using two additional graphs that mix the $\pi$ and $\sigma$ bonds equipped with additional correlators similar to Fig.~\ref{fig:systematic_errors_h4_c4h6_beh2}). This is however far from a well defined procedure and at this point not suitable for automatization which is why N$_2$/STO-3G(6,12) is considered a challenging system in the scope of this work. It is reasonable to assume that this will extend to other systems with triple-bonds and more challenging cases like the hextuple bond of the chromium dimer~\cite{larsson2022chromium}. Combined with circuits obtained via global-optimization of UCC type operators~\cite{burton2022exact} -- recently applied to N$_2$ -- better heuristics might be discovered in the future.\\
Although well-suited for state-of-the-art research in variational quantum circuit design the techniques described in this work are currently restricted to neutral molecules with even numbers of electrons. Extensions that go beyond this restricted formulation can however be envisioned. One way forward could be through ``restriced-open'' SPAs, containing a single unpaired electron, in their formulation as starting states. In appendix~\ref{sec:unpaired-electrons} we give an explicit example.

\section{Computational Details \& Code Availability}
The examples in the last section were computed using \textsc{tequila}~\cite{tequila,kottmann2020tequila} with optimized gradients and quantum chemistry framework described in Ref.~\cite{kottmann2021feasible}. We used \textsc{pyscf}~\cite{pyscf1,pyscf2} as backend for Gaussian integral evaluation and orbital optimization with more details on the latter described in Ref.~\cite{kottmann2022optimized}. Direct basis determination via MRA-PNOs was performed with \textsc{madness}~\cite{harrison2016madness}~\footnote{In particular we used the fork at \href{https://github.com/kottmanj/madness}{github.com/kottmanj/madness}. Integration into the main repository is planned.
} as described in Ref.~\cite{kottmann2020reducing} with the diagonal approximation described in Ref.~\cite{kottmann2022optimized} and the MP2-PNO implementation of Ref.~\cite{kottmann2020direct} without cusp regularization. The latter interfaces mean-field implementations described in~\cite{harrison2004multiresolution, bischoff2014regularizing}. Quantum circuits were processed and simulated automatically through \textsc{tequila} using \textsc{qulacs}~\cite{qulacs} as backend and the Jordan-Wigner encoding implemented within \textsc{openfermion}~\cite{OpenFermion}. Optimization of the circuit parameters was performed with the automatically differentiable framework~\cite{kottmann2021feasible} within \textsc{tequila} using \textsc{jax}~\cite{jax} and \textsc{scipy}~\cite{scipy}.
The \textsc{tequila} library is available on github and an explicit code example is provided in the appendix.

\section{Conclusion \& Outlook}
In this work molecular circuit design heuristics based on chemical graphs were developed. The heuristics show a good balance between hardware-efficiency and physically motivated design principles by producing relatively shallow and low-parametrized circuits with good convergence properties. This provides a way forward in all mayor challenges of variational quantum circuit design: Initial state generation by reducing the first graph to a classically tractable separable pair approximation. This is furthermore used to get an improved set of orbitals (compared to standard Hartree-Fock).
Operator ordering and parameter initialization is achieved through chemical insight guided by the structure of the chemical graphs.\\
So far, the molecular circuits were constructed manually by selecting graphs, orbital rotators and the corresponding initial orbital guesses through chemical intuition. The construction therefore still contains a relatively high human factor and is potentially challenging without some background in electronic structure. 
The original work on SPA circuits~\cite{kottmann2022optimized} used directly determined pair-natural orbitals~\cite{kottmann2020direct} that can be computed through a black-box procedure without background knowledge. 
Equipped with further heuristics, the techniques from this work could lead to an automatized construction of molecular quantum circuits and orbitals in a similar manner. In combination with the basis-set-free framework of~\cite{kottmann2020reducing}, existing automatization protocols~\cite{stein2016automated}, and explicitly correlated corrections~\cite{schleich2021improving} a path towards an automatized black-box approach with accurate numerical precision could be envisioned. In particular, future heuristics could benefit from modern correlation measures~\cite{ding2022quantum,boguslawski2013orbital,krumnow2016fermionic} in order to effectively place electronic correlators in the spirit of~\cite{zhang2020mutual}. An interesting direction towards improved heuristics could also be towards machine learning approaches, as they are for example developed to generate molecular graphs with specific properties~\cite{krenn2020self}, in order to generate and rank important graphs for a given set of molecular coordinates as well as to generate suitable initial guesses for orbital rotations. The combination with recent developments of circuit re-compilation~\cite{meister2022exploring, gustiani2022exploiting} offers an interesting path towards hardware-adapted circuits with improved properties.

\section*{Acknowledgement}
I would like to thank Philipp Schleich, Abhinav Anand, and Korbinian Kottmann for proofreading and feedback on the manuscript. In addition I thank Mario Krenn, Rodrigo A. Vargas-Hernandez, Alba Cervera-Lierta, Korbinian Kottmann and Al\'an Aspuru-Guzik for helpful discussions in the past that helped me find the inspiration for this work. 

\bibliographystyle{unsrtnat}
\bibliography{main}
\clearpage

\appendix
\section{Molecular Hamiltonians}\label{sec:molecular_hamiltonian}
Molecular Hamiltonians are predominantly represented in second quantization
\begin{align}
    H = \sum_{kl} h_{kl} a_k^\dagger a_l + \sum_{klmn} g_{klmn} a^\dagger a^\dagger a_n a_m\label{eq:molecular_hamiltonian}
\end{align}
with fermionic creation (annihilation) operators $a^\dagger_k$ ($a_k$) that create (annihilate) electrons in spin orbitals $\phi_{k\sigma}$ with $\sigma=\uparrow$ for even $k$ and $\sigma=\downarrow$ for odd $k$) and coefficients $h_{kl}$ and $g_{klmn}$ that denote integrals over the corresponding real-space operators (kinetic and potential energy result in $h_{kl}$, two-electron repulsion energy results in $g_{klmn}$) and spatial (\textit{i.e.} 3-dimensional) basis functions, called orbitals, $\phi_k(x,y,z)$. Note, that the choice of the basis orbitals $\phi_k$ is not unique and in principle \textit{any} set of orthonormal functions with reasonable properties can be used with the most prominent choice being atomic orbitals (predefined atom-centered sets of functions that resemble the solution to the hydrogen atom -- see~\eqref{eq:h2_atomics_cartoon} for a graphical illustration) For more details on basis function see ~\cite{kottmann2020reducing} (first paragraph of I for a short summary) and Sec.~VI.A.2 of~\cite{bharti2022noisy}. In this work, orthonormalized atomic basis functions from the STO-3G set are mostly used. This choice was made for simplicity.\\
Qubit Hamiltonians of molecules are usually constructed by encoding a second-quantized fermionic Hamiltonian in the form of~\eqref{eq:molecular_hamiltonian} to qubits. In this work the Jordan-Wigner transformation is used
\begin{align}
    a_i &\xrightarrow{JW} \sigma^+_i \prod_{k<i} \sigma_z^k\\
    a_i^\dagger &\xrightarrow{JW} \sigma^-_i \prod_{k<i} \sigma_z^k
\end{align}
with $\sigma_{\pm} = \frac{1}{2}\lr{\sigma_x \pm i\sigma_y}$. Other encodings~\cite{bravyi2002fermionic, chien2020custom, setia2018bravyikitaevsuperfast, derby2021compact} could however be used as well. Note however, that the explicit form of the pair-restricted two-electron correlator in~\eqref{eq:h4_u1c} and the individual Pauli terms in~\eqref{eq:ur_paulis} will differ.

\section{Orbital Rotation Circuits}\label{sec:orbital_rotation_circuits}
Here we give the graphical decomposition of orbital rotations as given in~\eqref{eq:h2_R}. In second quantization an orbital rotation between spin-orbitals $\phi_{i_\uparrow}$ and $\phi_{j_\uparrow}$ are defined as
\begin{align}
    U_{i_\uparrow}^{j_\uparrow} = e^{-i\frac{\theta}{2} \lr{a^\dagger_{i_\uparrow}a_{j_\uparrow} - a^\dagger_{j_\uparrow}a_{i_\uparrow}}} 
\end{align}

\begin{align}
    \lr{U_{i_\uparrow}^{j_\uparrow}}^\dagger a_i^\dagger U_{i_\uparrow}^{j_\uparrow} = \cos\lr{\frac{\theta}{2}} a_{i_\uparrow} + \sin\lr{\frac{\theta}{2}} a_{j_\uparrow}
\end{align}
which corresponds to transforming the orbitals themselves as $\phi_{i_\uparrow} + \phi_{j_\uparrow}$. An important consequence is, that the structure of the fermionic Hamiltonian (and in the same manner the encoded qubit Hamiltonian) is preserved under those transformations. In particular, transformed orbitals $\tilde{\phi}_i = \sum_k c_{ik} \phi_k$ will lead to a Hamiltonian as in~\eqref{eq:molecular_hamiltonian} with transformed coefficients (molecular integrals) like $\tilde{h}_{kl} = c_{km}c_{ln} h_{mn}$. Trailing orbital rotations on a quantum circuit can therefore always be absorbed into the Hamiltonian by transforming the corresponding coefficients in the fermionic Hamiltonian accordingly (see Ref.~\cite{schleich2023partitioning} for applications). As a consequence, the trailing operations in the circuits of the article (e.g. in Fig.~\ref{fig:overview_cartoon}, and in \ref{eq:h6_spa+}) do not need to be applied as quantum gates necessarily. In Heuristic~\ref{heuristic:SPA} this is already used by directly solving for the optimized orbital rotation angles $\theta$ through a second order expansion of $U^\dagger H U$ where U contains all orbital rotations. We refer to Ref.~\cite{kottmann2022optimized} for more details on the implementation used in this work and to Refs.~\cite{sokolov2020quantum, mizukami2020, yalouz2021stateaveraged} (in particular the appendix of~\cite{sokolov2020quantum}) for other examples of orbital-optimized circuits using the same technique.\\

In Jordan-Wigner encoding the two orbitals are mapped to qubits $2i$ and $2j$ and there spin-down counterparts to qubits $2i+1$ and $2j+1$ respectively.
An individual spin-orbital rotation between orbitals $\phi_{0_\uparrow}$ (mapped to qubit 0) and $\phi_{1\uparrow}$ (mapped to qubit 2) in Jordan-Wigner representation takes the form of two three-qubit Pauli rotations
\begin{align}
e^{-\frac{\theta}{2} \sigma^x_0 \sigma^z_1 \sigma^y_2} e^{-\frac{\theta}{2} \sigma^y_0 \sigma^z_1 \sigma^x_2}\label{eq:ur_paulis}
\end{align}
which can be graphically represented by 
\begin{align}
    \includegraphics[height=0.1\textwidth]{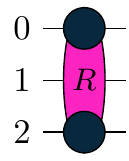}
    \raisebox{0.5cm}{$=$}
    \includegraphics[height=0.1\textwidth]{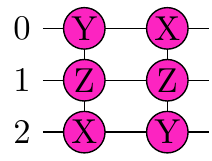}
\end{align}
and where each multi-Pauli rotation can be decomposed like
\begin{align}
    \includegraphics[height=0.1\textwidth]{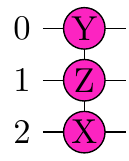}
    \raisebox{0.5cm}{$=$}
    \includegraphics[height=0.1\textwidth]{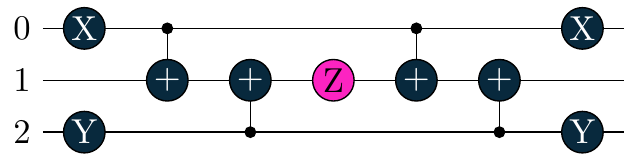}
\end{align}
where the first X and Y rotations have static angles of $\frac{\pi}{2}$ and the trailing ones $-\frac{\pi}{2}$. For more details on the decomposition of Pauli rotations into standard gates see the review~\cite{anand2022quantum} and Ref.~\cite{tequila} for the corresponding implementation within the \textsc{tequila} framework.\\

In the same manner, the spatial orbitals can be transformed by applying the spin-up and spin-down transformations in sequence $U_i^j = U_{i_\uparrow}^{j_\uparrow }U_{i_\downarrow}^{j_\downarrow}$ with their respective angles being identical.\\
\section{Details on the frozen-core errors of LiH}\label{sec:lih_details}
\begin{figure}
    \centering
    \includegraphics[width=0.75\textwidth]{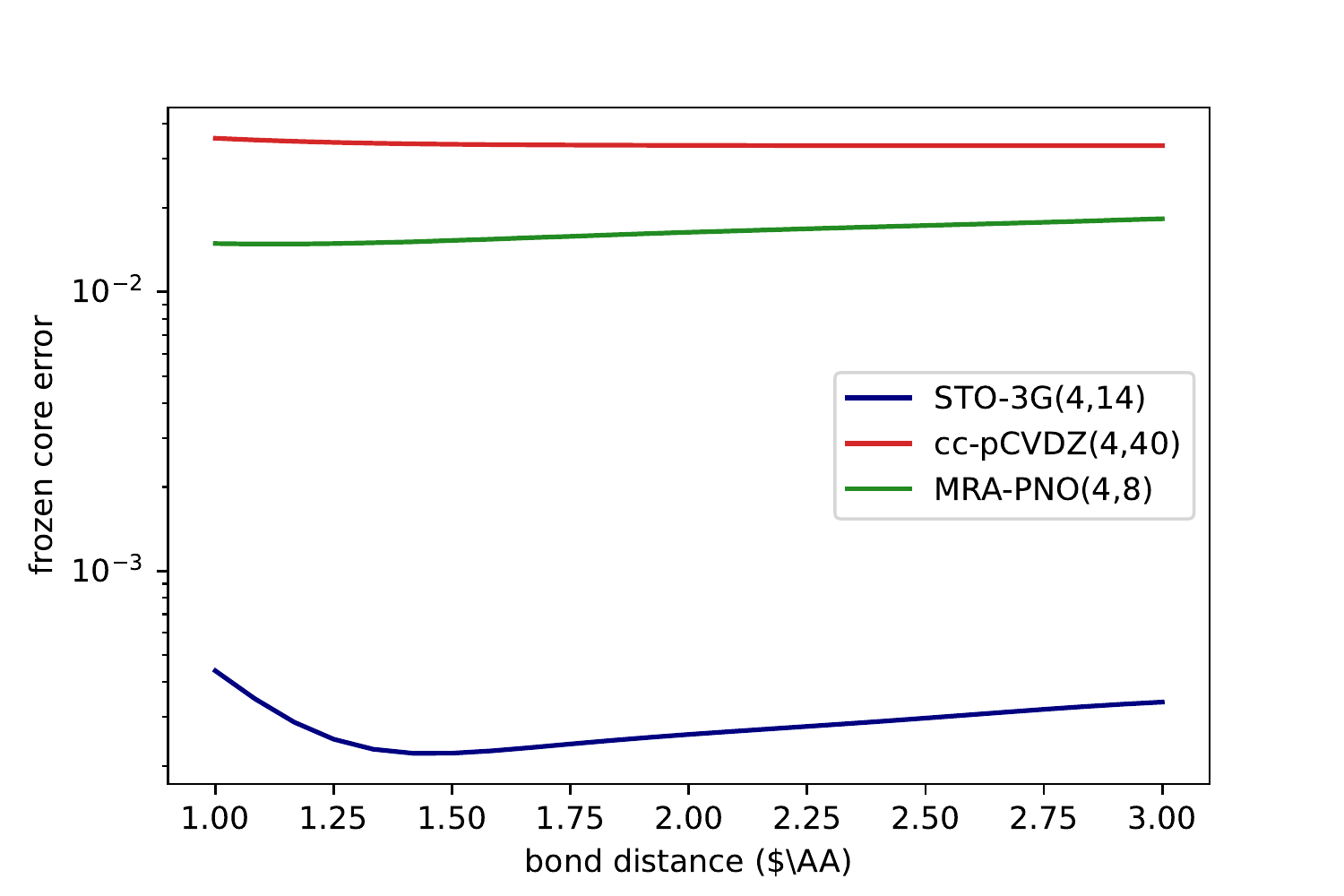}
    \caption{Frozen-core errors for LiH with different basis sets. The errors are computed as $|FCI/basis(4,X) - FCI/basis(2,X-2)|$ with the lowest Hartree-Fock orbital frozen.}
    \label{fig:lih_frozen_core_errors}
\end{figure}
In Fig.~\ref{fig:lih_frozen_core_errors} we show the frozen-core errors of LiH in the minimal basis STO-3G, a larger basis with specialized core-correlation functions cc-pCVDZ and a customized MRA-PNO basis that was constructed to include correlation in the core electrons (by optimizing one function for each PNO-MP2 pair~\cite{kottmann2020direct, kottmann2020reducing, kottmann2022optimized}). The STO-3G basis is missing \textit{core polarization} functions, so that there is no notable representation of that correlation in the basis, leading to a frozen-core error below the millihartree threshold for all bond distances. Here the frozen-core errors is identical to the SPA/STO-3G(2,12) error in Fig.~\ref{fig:h2_lih_results} of the main text. The cc-pCVDZ basis admids a significant frozen-core error, due to the presence of specialized core-polarization functions in the basis. The cc-pCVDZ basis is inconvenient for the investigation of the SPA performance as it would require 40 qubits to represent. This is why we employed a system adapted MRA-PNO basis in Fig.~\ref{fig:h2_lih_results} in order to demonstrate that the LiH molecule can be efficiently represented by separable pairs. As the frozen-core error for the MRA-PNOs is significant and comparable to cc-pCVDZ, the SPA wavefunctions includes two electrons pairs. The opt-SPA results is again below the millihartree threshold for all bond distances. This shows, that, despite having significant core correlation in the system, the corresponding wavefunction is still separable into two electron pairs. In chemical terms this means, that the core electrons are not correlated with the valence (``bonding'') electrons in a significant way. 

\section{Optimized Orbitals}\label{sec:optimized_orbitals}
In the article we represent orbitals with small cartoons that depict them as linear combinations of their respective atomic basis orbitals. The optimized orbitals for H2/STO-3G(2,4) were for example shown as
\begin{align}
    \includegraphics[width=0.05\textwidth, angle=90]{h2_atomic_orbitals.pdf}\;\raisebox{0.3cm}{$\xrightarrow{U_{\text{R}}}$}\;
   \includegraphics[width=0.05\textwidth]{h2_hf_orbitals.pdf},
\end{align}
with the two atomic orbitals on the left, and the two optimized combination son the right.
The corresponding matrix, holding the coefficients that define the optimized orbitals in a linear superposition of atomic basis orbitals is
\begin{align}
    \begin{pmatrix}
    \frac{1}{\sqrt{2}} & \frac{1}{\sqrt{2}}\\
    \frac{1}{\sqrt{2}} & -\frac{1}{\sqrt{2}}\\
    \end{pmatrix},
\end{align}
and $U_\text{R}$ represents the corresponding quantum circuit~\eqref{eq:h2_R} that has the same effect (see Sec~\ref{sec:orbital_rotation_circuits}) when acting on an Hamiltonian $H$ formed from orthonormalized atomic basis orbitals -- \textit{i.e.} $H'=U_\text{R}^\dagger HU_\text{R}$ is the same as a qubit Hamiltonian formed from orbitals obtained with the matrix above.
In the following the optimized coefficients with respect to Graph-SPA circuits are displayed as linear combinations of their respective (non-normalized) atomic orbitals of the corresponding STO-3G basis set.
\subsection{Optimized H$_4$ Orbitals}
Coefficient matrix for the optimized orbitals of the linear H$_4$ system represented in \eqref{eq:h4_optimized_orbitals_cartoon}:
\begin{align}
\begin{pmatrix}
 +0.7044 & +0.7003 & +0.0824 & -0.0819 \\
 +0.7003 & -0.7044 & +0.0819 & +0.0824 \\
 -0.0819 & +0.0824 & +0.7003 & +0.7044 \\
 -0.0824 & -0.0819 & +0.7044 & -0.7003 \\
\end{pmatrix}.
\end{align}
Rows denote the coefficients of the optimized orbitals and columns represent the s orbitals from the STO-3G set (from left to right in the linear chain). See the code example in~\ref{sec:code_example_h4} on how the coefficients can be computed.
\subsection{Optimized BeH$_2$ Orbitals}
The coefficient matrix for BeH$_2$/STO-3G(4,8) with be-H inter-atomic distances 1.5{\AA} (close to the equilibrium):
\begin{align}
\begin{pmatrix}
 +0.9916 & +0.0322 & +0.0000 & -0.0034 & -0.0034 \\
 -0.1519 & +0.4168 & +0.3535 & +0.5775 & -0.0180 \\
 -0.1620 & +0.7782 & +0.8984 & -1.1509 & +0.0917 \\
 -0.1519 & +0.4167 & -0.3536 & -0.0181 & +0.5775 \\
 -0.1620 & +0.7782 & -0.8983 & +0.0916 & -1.1509 \\
\end{pmatrix}.
\end{align}
and with 6{\AA}~Be-H distances (effectively a dissociated system),
\begin{align}
    \begin{pmatrix}
 +0.9969 & +0.0117 & -0.0000 & -0.0000 & -0.0000 &\\
 -0.2800 & +1.0354 & +0.0000 & +0.0001 & +0.0001 &\\
 +0.0000 & -0.0000 & +1.0000 & -0.0004 & +0.0004 &\\
 +0.0001 & -0.0003 & +0.0000 & +0.7071 & +0.7071 &\\
 -0.0000 & +0.0000 & +0.0003 & +0.7071 & -0.7071 &\\
    \end{pmatrix}.
\end{align}
Rows represent the 5 optimized orbitals with the first row beeing the frozen core orbital. The columns denote the atomic orbitals from the STO-3G set in the order: Be-S, Be-S, Be-P$_\text{z}$, right-H-S, left-H-S. The initial guess for the optimization where equal weighted superpositions depicted in~\eqref{eq:orbitals_beh2} with no contribution of the first Be-S. The initial guess of the core orbital (not depicted in~\eqref{eq:orbitals_beh2} was only the first Be-S orbital.
\subsection{Optimized C$_4$H$_6$ Orbitals}
The coefficient matrix for the planar C$_4$H$_6$/STO-3G(4,8) close to equilibrium distance are
\begin{align}
\begin{pmatrix}
 +0.6352 & +0.6291 & +0.0342 & -0.0913 &\\
 +0.8007 & -0.8222 & +0.1483 & +0.0718 &\\
 -0.0913 & +0.0342 & +0.6291 & +0.6352 &\\
 +0.0718 & +0.1483 & -0.8222 & +0.8007 &\\
\end{pmatrix}.
\end{align}
Rows represent the 4 optimized $\pi$-orbitals with and columns denote the atomic P$_z$ orbitals from the STO-3G from left to right in the linear chain. The nuclear coordinates (x,y,z) of the molecule are 
\begin{align}
    \begin{pmatrix}
    C\\C\\C\\C\\H\\H\\H\\H\\H\\H\\
    \end{pmatrix}
    \rightarrow
    \begin{pmatrix}
+1.1216080&+1.4917400&+0.0\\
+0.0000000&+0.7410470&+0.0\\
+0.0000000&-0.7410470&+0.0\\
-1.1216080&-1.4917400&+0.0\\
-0.9808550&+1.2258790&+0.0\\
+0.9808550&-1.2258790&+0.0\\
+1.0877710&+2.5803400&+0.0\\
+2.1146690&+1.0387500&+0.0\\
-1.0877710&-2.5803400&+0.0\\
-2.1146690&-1.0387500&+0.0\\
    \end{pmatrix}
\end{align}

\section{Full H$_6$ Circuit}
The SPA+ circuit for the linear H$_6$/STO-3G(6,12) system is
\begin{align}
        \includegraphics[height=0.5\textwidth]{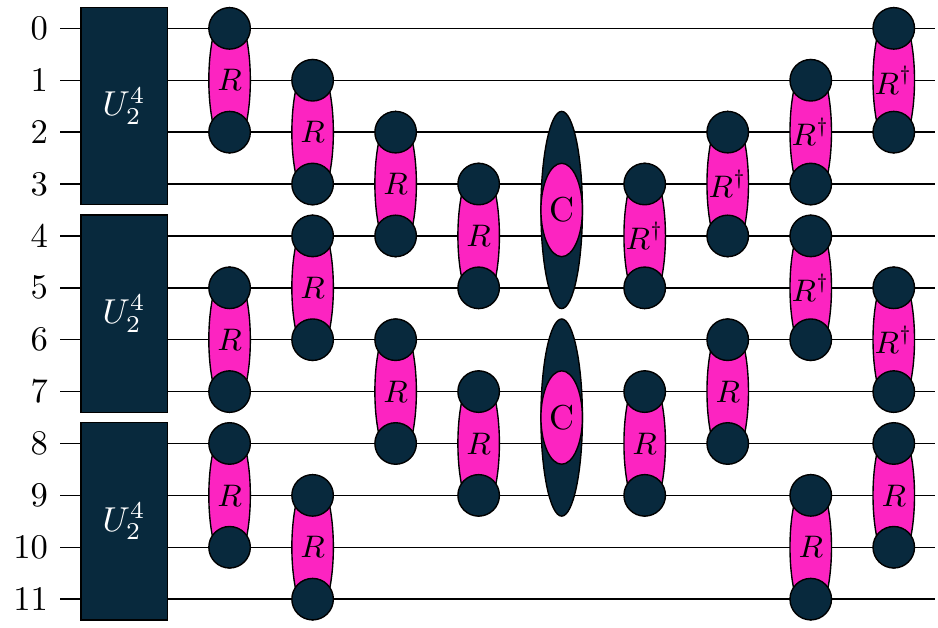}\label{eq:h6_spa+}
\end{align}
with $U_2^4$ blocks defined in~\eqref{eq:h2_spa_circuit} and the molecular correlators C, and orbital rotations R defined according to~\eqref{eq:h4_u1c} and ~\eqref{eq:h2_R}. 

\clearpage
\section{Code Example: H$_4$}\label{sec:code_example_h4}
\begin{lstlisting}[language=Python]
# dependencies: tequila >= 1.8.7, pyscf~=1.7, scipy~=1.7
# suggested quantum backend for optimal performance: qulacs >= 0.3
import tequila as tq
from numpy import eye, pi

# Create the molecule object
# use orthonormalized atomic orbitals as basis
geometry = "h 0.0 0.0 0.0\nh 0.0 0.0 1.5\nh 0.0 0.0 3.0\nh 0.0 0.0 4.5"
mol = tq.Molecule(geometry=geometry, basis_set="sto-3g")
energies = {"FCI":mol.compute_energy("fci")}
# switch from canonical HF orbitals to orthonormalized STO-3G orbitals
# to follow notation in the article
mol = mol.use_native_orbitals()

# Create the SPA circuit for
# Graph: H -- H    H -- H
# edges get tuples of orbital-indices assigned
USPA = mol.make_ansatz(name="SPA", edges=[(0,1),(2,3)])

# initial guess for the orbitals
# according to graph in Eq.(17) and orbitals in Eq.(19)
guess = eye(4)
guess[0] = [1.0,1.0,0.0,0.0]
guess[1] = [1.0,-1.,0.0,0.0]
guess[2] = [0.0,0.0,1.0,1.0]
guess[3] = [0.0,0.0,1.0,-1.]

# optimize orbitals and circuit parameter
# PySCF interface
opt = tq.chemistry.optimize_orbitals(mol, circuit=USPA, initial_guess=guess.T)
print("Optimized Orbital Coefficients")
print(opt.molecule.integral_manager.orbital_coefficients)
energies["SPA"] = opt.energy

# get Hamiltonian with optimized orbitals
H = opt.molecule.make_hamiltonian()

# initialize rotations to graph in Eq.(21)
# H    H -- H    H
# as illustrated in Eq.(24)
# UR as in Eq.(7) uses spatial orbital indices
R0 = tq.Variable("R0")
R1 = tq.Variable("R1")
UR0 = mol.UR(0,1,angle=(R0+0.5)*pi)
UR0+= mol.UR(2,3,angle=(R0+0.5)*pi)
UR1 = mol.UR(1,2,angle=(R1+0.5)*pi)

# initialize correlator according to Eq.(22)
UC = mol.UC(1,2,angle="C")

# construct the circuit for both graphs
U = USPA + UR0 + UR1 + UC + UR1.dagger() + UR0.dagger()
# optimize the energy
E = tq.ExpectationValue(H=H, U=U)
result = tq.minimize(E, silent=True)
energies["SPA+"] = result.energy
print(energies)
\end{lstlisting}

\section{Comment on systems with odd electron number}\label{sec:unpaired-electrons}

The main article focuses primarily on systems with even electron number, which are currently the dominating usecases in comparable literature. Extensions of the current framework to odd-numbered electrons is possible, but will require some work in the underlying details. In the following a sketch for a potential approach will be given, using the neutral H$_3$/STO-3G(3,6) molecule -- a 3-electron system in 6 qubits -- as example.\\

This work used SPA circuits in order to represent the first molecular graph. SPAs were developed for even-numbered electrons as well, and although we could extend them to odd-numbered electrons by simply replacing one of the electron-pairs with a single electron, we will chose a simpler strategy where all graphs are represented with the same $U_\text{R}U_\text{C}U_\text{R}$ motif. As an initial state (which in the main article is included in the SPA construction), we then chose a basis state with the right amount of electrons. In the H$_3$ case this can for example be
\begin{align}
    \ket{\Psi_0} = \ket{110010},
\end{align}
with two spin-paired electrons in the first, and a single electron in the third orbital. On this initial state we are then applying unitaries that correspond to specific molecular graphs (as in the main text)
\begin{align}
    \ket{\Psi(\boldsymbol{a},\boldsymbol{b},\boldsymbol{c}} = U(\boldsymbol{c})U(\boldsymbol{b})U(\boldsymbol{a})\ket{\Psi_0} \label{eq:h3_circuit}
\end{align}
with the graphs being
\begin{align}
    \includegraphics[width=0.75\textwidth]{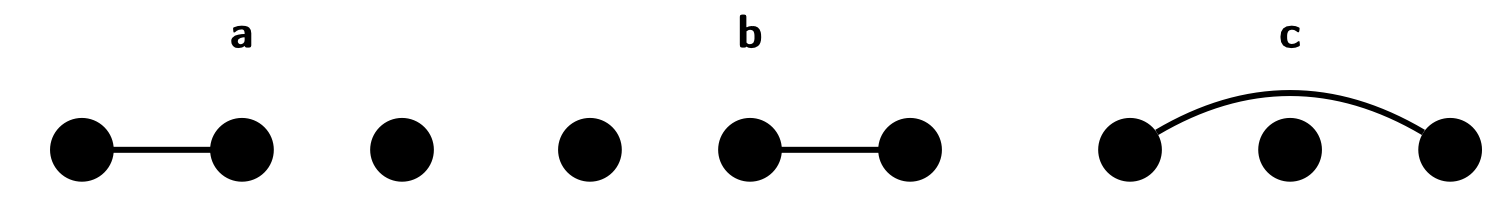}
\end{align}
In Tab.~\ref{tab:h3results} we show the performance of such circuits where we, as in the main article, started from atomic orbitals and either used a single orbital rotator per graph (corresponding to a localized orbital picture) or allowed more freedom by placing three rotators per graph (corresponding to full orbital optimization). Similar to the H$_4$ results in the main text we see that graphs $\textbf{a}$ and $\textbf{b}$ are dominant and lead to exact results when the circuits are equipped with enough freedom in the orbital rotations.

\begin{table}
    \centering
    \begin{tabular}{ccc}
    \toprule
    \# Graphs & \multicolumn{2}{c}{ $U_\text{R}$ per Graph} \\
     & 1 & 3 \\ 
    \midrule
       1  & 55 & 53\\
       2  & 33 & 0\\
       3  & 24 & 0 \\
    \midrule
    \bottomrule
    \end{tabular}
    \caption{Molecular Circuits for H$_3$: Energy errors with respect to exact diagonalization in millihartree. Energies are obtained with circuits constructed according to Eq.~\eqref{eq:h3_circuit}. Where the individual unitaries are either implementing a single or three orbital rotators per graph.}
    \label{tab:h3results}
\end{table}

\end{document}